\documentclass[journal]{IEEEtran}

\ifCLASSINFOpdf
\else
\fi
\usepackage[dvips]{graphicx}
\usepackage{latexsym}
\usepackage{amssymb}
\usepackage{amsmath}
\usepackage{multirow}
\usepackage{xcolor}
\usepackage{lipsum}
\usepackage{multicol}
\usepackage{enumerate}
\usepackage{algorithm}
\usepackage{algorithmicx}
\usepackage{algpseudocode}
\usepackage{amsmath}
\usepackage{graphicx}
\usepackage{subfig}
\usepackage[utf8]{inputenc}
\usepackage{float}

\begin{document}
%
% paper title
% can use linebreaks \\ within to get better formatting as desired
% Do not put math or special symbols in the title.
\title{Massive MIMO 1-Bit DAC Transmission: A Low-Complexity Symbol Scaling Approach}
%
%
% author names and IEEE memberships
% note positions of commas and nonbreaking spaces ( ~ ) LaTeX will not break
% a structure at a ~ so this keeps an author's name from being broken across
% two lines.
% use \thanks{} to gain access to the first footnote area
% a separate \thanks must be used for each paragraph as LaTeX2e's \thanks
% was not built to handle multiple paragraphs
%

\author{Ang~Li,~\IEEEmembership{Student Member,~IEEE}, Christos Masouros,~\IEEEmembership{Senior Member,~IEEE}, Fan Liu,~\IEEEmembership{Student Member,~IEEE}, and A. Lee Swindlehurst,~\IEEEmembership{Fellow,~IEEE}

        % <-this % stops a space
% <-this % stops a space <-this % stops a space
%\thanks{Manuscript received TIME; revised TIME.}
\thanks{A. Li and C. Masouros are with the Department of Electronic and Electrical Engineering, University College London, Torrington Place, London, WC1E 7JE, UK (e-mail: c.masouros@ucl.ac.uk, ang.li.14@ucl.ac.uk).}
\thanks{F. Liu is with the Department of Electronic and Electrical Engineering, University College London, Torrington Place, London, WC1E 7JE, UK, and also with the School of Information and Electronics, Beijing Institute of Technology, Beijing 100081, China (e-mail: liufan92@bit.edu.cn).}
\thanks{A. L. Swindlehurst is with the Department of Electrical Engineering and Computer Science, Henry Samueli School of Engineering, University of California, Irvine, CA 92697 USA, and also with the Institute for Advanced Study, Technical University of Munich, 80333 Munich, Germany (e-mail: swindle@uci.edu).}
\thanks{This work was supported by the Royal Academy of Engineering, UK, the Engineering and Physical Sciences Research Council (EPSRC) project EP/M014150/1 and the China Scholarship Council (CSC).}
}

% If you want to put a publisher's ID mark on the page you can do it like
% this:
%\IEEEpubid{0000--0000/00\$00.00~\copyright~2012 IEEE}
% Remember, if you use this you must call \IEEEpubidadjcol in the second
% column for its text to clear the IEEEpubid mark.

% use for special paper notices
%\IEEEspecialpapernotice{(Invited Paper)}

% make the title area
\maketitle

% As a general rule, do not put math, special symbols or citations
% in the abstract or keywords.
\begin{abstract}
We study multi-user massive multiple-input single-output (MISO) systems and focus on downlink transmission, where the base station (BS) employs a large antenna array with low-cost 1-bit digital-to-analog converters (DACs). The direct combination of existing beamforming schemes with 1-bit DACs is shown to lead to an error floor at medium-to-high SNR regime, due to the coarse quantization of the DACs with limited precision. In this paper, based on the constructive interference we consider both a quantized linear beamforming scheme where we analytically obtain the optimal beamforming matrix, and a non-linear mapping scheme where we directly design the transmit signal vector. Due to the 1-bit quantization, the formulated optimization for the non-linear mapping scheme is shown to be non-convex. To solve this problem, the non-convex constraints of the 1-bit DACs are firstly relaxed, followed by an element-wise normalization to satisfy the 1-bit DAC transmission. We further propose a low-complexity symbol scaling scheme that consists of three stages, in which the quantized transmit signal on each antenna element is selected sequentially. Numerical results show that the proposed symbol scaling scheme achieves a comparable performance to the optimization-based non-linear mapping approach, while its corresponding complexity is negligible compared to that of the non-linear scheme.
\end{abstract}

% Note that keywords are not normally used for peerreview papers.
\begin{IEEEkeywords}
Massive MIMO, 1-bit quantization, beamforming, constructive interference, Lagrangian, low-complexity scheme.
\end{IEEEkeywords}

% For peer review papers, you can put extra information on the cover
% page as needed:
% \ifCLASSOPTIONpeerreview
% \begin{center} \bfseries EDICS Category: 3-BBND \end{center}
% \fi
%
% For peerreview papers, this IEEEtran command inserts a page break and
% creates the second title. It will be ignored for other modes.
\IEEEpeerreviewmaketitle

\section{Introduction}
% The very first letter is a 2 line initial drop letter followed
% by the rest of the first word in caps.
%
% form to use if the first word consists of a single letter:
% \IEEEPARstart{A}{demo} file is ....
%
% form to use if you need the single drop letter followed by
% normal text (unknown if ever used by IEEE):
% \IEEEPARstart{A}{}demo file is ....
%
% Some journals put the first two words in caps:
% \IEEEPARstart{T}{his demo} file is ....
%
% Here we have the typical use of a "T" for an initial drop letter
% and "HIS" in caps to complete the first word.

\IEEEPARstart{T}{OWARDS} the fifth generation (5G) and future wireless communication systems, massive multiple-input multiple-output (MIMO) systems \cite{r10} have received increasing research attention in recent years as they are able to greatly improve the spectral efficiency. It has also been shown that low-complexity linear precoding approaches such as zero-forcing (ZF) \cite{r5} and regularized ZF (RZF) \cite{r6} achieve close-to-optimal performance in the massive MIMO regime. Nevertheless, with a large number of antennas employed at the BS, the large number of radio frequency (RF) chains and corresponding digital-to-analog converters (DACs) that need to be employed at the BS pose a significant practical challenge. This increase in the hardware complexity and resulting power consumption hinders the practical implementation of massive MIMO. To achieve a compromise between the performance, hardware complexity and the consequent power consumption in practical massive MIMO systems, hybrid analog digital beamforming \cite{r40}, \cite{r41} has attracted research interest as a means of reducing the number of RF chains. 

In addition to the hybrid structures, another potential approach, which is the focus of this paper, is to reduce the cost and power consumption per RF chain by employing very low-resolution digital-to-analog converters (DACs) instead of high-precision DACs. It has been shown in \cite{r13} that DACs are one of the dominant power-consuming hardware components in the downlink, whose power consumption grows exponentially with the resolution and linearly with the bandwidth. In the traditional MIMO downlink, each transmit signal is generated by a pair of high-resolution (usually more than 8-bit) DACs that are connected to the RF chain. However, in the case of massive MIMO with hundreds of antennas employed at the BS, a large number of DACs are required and the resulting power consumption will be prohibitively high. Therefore, employing low-resolution DACs, especially 1-bit DACs, can greatly reduce the power consumption per RF chain and the resulting total power consumed at the BS. When 1-bit DACs are employed, the output signal at each antenna element is equivalent to the constant-envelope symbol from a QPSK constellation, which enables the use of low-cost power amplifiers (PAs) and can further reduce the hardware complexity. 

In the existing literature, most recent studies have focused on the performance analysis for massive MIMO uplink with low-resolution analog-to-digital converters (ADCs), especially for the 1-bit case \cite{r14}\nocite{r15}-\cite{r16}, where it is shown that the number of quantization bits can be reduced while a comparable performance is still achievable. For the case of downlink transmission with 1-bit DACs, there have been an increasing number of studies due to the benefits mentioned above \cite{r17}\nocite{r18}\nocite{r19}\nocite{r38}-\cite{r20}. In \cite{r17}, a simple quantized ZF scheme is considered, where the transmit signal vector is obtained by a direct quantization on the ZF-precoded signals. The authors further analyze the performance of the quantized ZF scheme, and show that it outperforms the maximum likelihood (ML) encoder in the low-to-medium SNR regime. In \cite{r18}, \cite{r19}, the quantized linear beamforming schemes based on minimum-mean squared error (MMSE) are proposed, whose performance is shown to be superior to the quantized ZF scheme in \cite{r17}. In \cite{r38}, a non-linear symbol perturbation technique is introduced in 1-bit massive MIMO downlink for QPSK modulation, while in \cite{r20} an iterative non-linear beamforming scheme is introduced via a biconvex relaxation approach, where the proposed scheme directly designs the transmit signal vector based on the MMSE criterion. Nevertheless, while operating on a symbol-by-symbol basis, these MMSE-based schemes may be sub-optimal as they ignore the fact that interference can be exploited on an instantaneous basis in \cite{r22}\nocite{r36}\nocite{r9}\nocite{r31}\nocite{r33}-\cite{r21}. Moreover, while there have been studies on the downlink beamforming schemes with 1-bit DACs, most of the these existing schemes either suffer a severe performance degradation in \cite{r17}-\cite{r19} compared to the unquantized case, or require sophisticated optimizations and iterative algorithms that are computationally inefficient \cite{r20}.

In this paper, we revisit the symbol-level operations required for massive MIMO downlink transmission with 1-bit DACs to exploit the formulation of  constructive interference. The symbol-by-symbol precoding operation allows us to observe the interference from an instantaneous point of view, and exploit it constructively \cite{r22}-\cite{r21}. We firstly consider a quantized linear beamforming scheme by constructing a beamforming matrix before quantization. Based on the concept of constructive interference, the optimization aims to maximize the distance between the received symbols and the detection thresholds. By mathematically analyzing the optimization problem with the Lagrangian approach, it is shown that the optimality is achieved by applying a strict phase rotation for the constructed problem in the case of massive MIMO. Due to the operation of the 1-bit quantization, the above quantized linear scheme is analytically shown to be equivalent to the quantized ZF scheme, which suffers an error floor at high SNR. To improve the performance, we then propose a non-linear mapping scheme where we directly design the quantized transmit signal vector. Nevertheless, due to the constraint on the output signals of 1-bit DACs, the resulting optimization problem is shown to be non-convex. To solve this problem, we firstly apply a relaxation on the mathematical constraint resulting from the use of 1-bit DACs, such that the optimization problem becomes convex. Then, we apply an element-wise normalization on the signal vector obtained from the relaxed optimization to meet the constraint on the output signals of 1-bit DACs.

Nevertheless, since the variable of the non-linear optimization approach is the transmit signal vector, whose dimension is equal to the number of transmit antennas, the computational complexity of the resulting optimization is high in the case of massive MIMO. Therefore, to enable the practical implementation of 1-bit DACs, we further propose a low-complexity symbol scaling scheme based on a coordinate transformation of the constructive interference problem, where we directly select the 1-bit DAC output for each antenna element on a sequential basis, and a relaxation-normalization process is therefore no longer needed. The proposed symbol scaling approach consists of three stages: an initialization stage where we decide the output signals for some antenna elements whose channel coefficients satisfy certain requirements, an allocation stage where we sequentially select the output signals for the residual antenna elements, and a refinement stage where we check whether the performance with the obtained signal vector can be further improved based on the greedy algorithm. Both the `Sum-Max' and the `Max-Min' criteria are considered in the allocation stage, and the output signal vector that returns the best performance is then obtained within the above two criteria. We further study the computational costs of the proposed optimization-based and symbol scaling schemes in terms of the floating operations required. Numerical results show that in the case of small-scale MIMO systems, the proposed symbol scaling scheme is shown to achieve the best performance. In the case of massive MIMO, the optimization-based non-linear scheme achieves an improved performance over existing schemes and better approaches the unquantized scheme, while the proposed symbol scaling scheme can achieve a comparable performance. In terms of the computational complexity, it is demonstrated that the complexity of the symbol scaling scheme is negligible compared to that of the non-linear mapping approach, while the performance of the symbol scaling scheme is superior to `Pokemon' when their computational costs are similar, which favours its usefulness in practice.

\begin{figure*}[!t]
\centering
\includegraphics[scale=0.45]{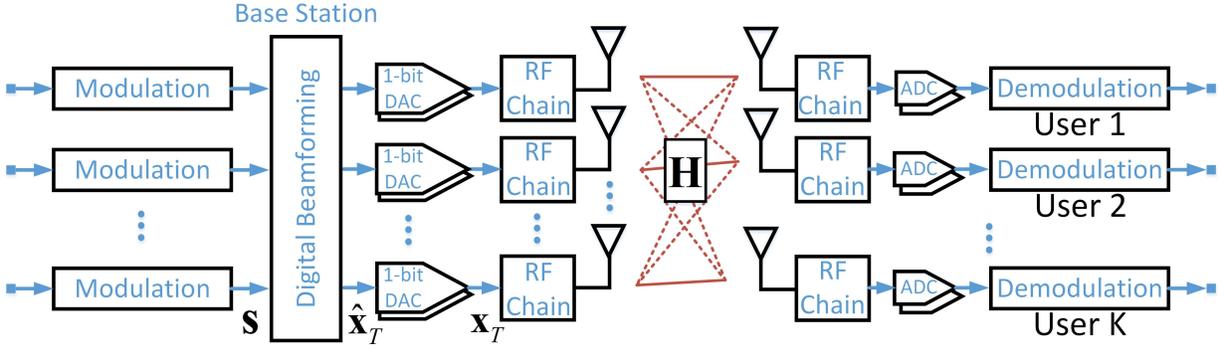}
\caption{Massive MIMO downlink system model with 1-bit DACs}
\end{figure*}

For reasons of clarity, we summarize the contributions of this paper as:
\begin{enumerate}

\item We propose downlink beamforming schemes for massive MIMO with 1-bit DACs based on the constructive interference formulation. We firstly consider a quantized linear beamforming scheme, where it is analytically proven that, in the massive MIMO region, the optimality is achieved by employing a strict phase rotation due to the favourable propagation conditions.

\item We then consider a non-linear mapping scheme where we directly optimize the transmit signal vector. The resulting non-convex optimization is solved in two steps: we firstly relax the non-convex constraints of 1-bit DACs, followed by the normalization on the obtained signal vector to satisfy the 1-bit DAC transmission.

\item Based on a coordinate transformation of the constructive interference formulation, we further propose a low-complexity symbol scaling scheme where we directly select the quantized signal on each antenna element via a three-stage process. It is shown that the symbol scaling scheme can achieve a comparable performance to the optimization-based non-linear mapping scheme.

\item We further study and compare the computational costs of the optimization-based non-linear mapping scheme and the symbol scaling schemes in terms of the floating operations required, where it is shown mathematically and numerically that compared to the non-linear mapping approach, the complexity of the proposed symbol scaling approach is negligible.

\end{enumerate}

The remainder of this paper is organized as follows. Section II introduces the system model. Both the proposed optimization-based quantized linear beamforming scheme and the non-linear mapping scheme that exploit the constructive interference are presented in Section III. The low-complexity three-stage symbol scaling method is presented in Section IV. Section V includes the analysis of the computational complexity for both schemes, and the numerical results are shown in Section VI. Section VII concludes the paper.

$Notations$: $a$, $\bf a$, and $\bf A$ denote scalar, vector and matrix, respectively. ${( \cdot )^T}$ and ${( \cdot )^H}$ denote transposition and conjugate transposition of a matrix, respectively. $card \left( {\cdot} \right)$ denotes the cardinality of a set. $j$ denotes the imaginary unit, and ${\rm{vec}}\left(  \cdot  \right)$ denotes the vectorization operation. ${\bf{a}}\left( k \right)$ denotes the $k$-th entry in vector $\bf a$. $\left|  \cdot  \right|$ denotes the modulus of a complex number or the absolute value of a real number, $\left\|  \cdot  \right\|_F$ denotes the Frobenius norm, and $\left\|  \cdot  \right\|_1$ denotes the 1-norm. ${{\cal C}^{n \times n}}$ represents an $n \times n$ matrix in the complex set, and $\bf I$ denotes the identity matrix. $\Re ( \cdot )$ and $\Im ( \cdot )$ denote the real and imaginary part of a complex number, respectively.

\section{System Model}
We consider a multi-user massive MIMO downlink, where 1-bit DACs are employed at the BS, as depicted in Fig. 1. As we focus on the transmit-side processing, ideal ADCs with infinite precision are assumed to be employed at each receiver. The BS with $N_t$ transmit antennas is communicating with $K$ single-antenna users simultaneously in the same time-frequency resource, where $K \ll {N_t}$. We focus on the transmit beamforming designs and perfect CSI is assumed, while we also numerically study the performance of the proposed schemes with imperfect CSI in Section VI. Following the closely-related literature \cite{r17}-\cite{r38}, \cite{r37}, the symbol vector is assumed to be from a normalized PSK constellation. We denote the data symbol vector as ${\bf{s}} \in {{\cal C}^{K \times 1}}$, and the unquantized signal vector that is formed based on $\bf s$ as ${{\bf{\hat x}}_T} \in {{\cal C}^{{N_t} \times 1}}$. Then, the unquantized signal vector ${\bf \hat x}_T$ can be expressed as
\begin{equation}
{{\bf{\hat x}}_T} = {\cal B}\left( {\bf{s}} \right),
\end{equation}
where $\cal B$ denotes a general linear or non-linear transformation. With 1-bit DACs employed, the output signal vector is then obtained as
\begin{equation}
{{\bf{x}}_T} = {\cal Q}\left( {{{{\bf{\hat x}}}_T}} \right).
\end{equation} 
In (2), $\cal Q$ denotes the 1-bit quantization on both the real and imaginary part of each entry in ${\bf \hat x}_T$. We denote $x_n$, $n \in \left\{ {1,2,\cdots,{N_t}} \right\}$ as the $n$-th entry in ${\bf x}_T$, and in this paper each $x_n$ is normalized to satisfy
\begin{equation}
{x_n} \in \left\{ { \pm \frac{1}{{\sqrt {2{N_t}} }} \pm \frac{1}{{\sqrt {2{N_t}} }} \cdot j} \right\}, {\kern 2pt} \forall n \in \cal N,
\end{equation}
where ${\cal N}=\left\{ {1,2,...,{N_t}} \right\}$. The above normalization guarantees that $\left\| {{{\bf{x}}_T}} \right\|_F^2 = 1$, and we can then express the received signal vector as
\begin{equation}
{\bf{y}} = \sqrt P  \cdot {\bf{H}}{{\bf{x}}_T} + {\bf{n}},
\end{equation}
where ${\bf H} \in {\cal C}^{K \times N_t}$ denotes the flat-fading Rayleigh channel with each entry following a standard complex Gaussian distribution. ${\bf n} \in {\cal C}^{K \times 1}$ denotes the additive Gaussian distributed noise vector with zero mean and covariance ${\sigma ^2} \cdot {\bf{I}}$. $P$ is the total available transmit power per antenna, and for simplicity in this paper we assume uniform power allocation for the antenna array.

\section{1-Bit Transmission Scheme based on Constructive Interference}
%In this section, we present the proposed optimization-based schemes that exploit the constructive interference. The concept of constructive interference and constructive region is first briefly reviewed, followed by the description of the proposed linear and non-linear 1-bit transmission schemes.

\subsection{Constructive Interference and Constructive Region}
Constructive interference is defined as interference that pushes the received signals away from the detection thresholds of the modulation constellation \cite{r22}-\cite{r9}. The exploitation of constructive interference was firstly introduced in \cite{r22} to improve the performance of the ZF beamforming scheme, and was more recently applied to optimization-based approaches in \cite{r36}, \cite{r9} and \cite{r21} based on the constructive region. To illustrate the underlying concept intuitively, in Fig. 2 we depict the constructive region for QPSK, where for simplicity and without loss of generality we focus on one quarter of the normalized QPSK constellation. As can be observed, as long as the interfered signal ($\mathop {OB}\limits^ \to$ in Fig. 2) is located in the constructive region, the distance to the detection thresholds is increased, and an improved performance can be expected. The formulation of the optimization problem based on the constructive region will be introduced in the following.

%Constructive interference is defined as interference that pushes the received signals away from the detection thresholds of the modulation constellation \cite{r9}. The exploitation of constructive interference was firstly introduced in \cite{r22} to improve the performance of the ZF beamforming scheme, and was more recently applied to optimization-based approaches \cite{r36}, \cite{r9} and \cite{r21}. It is demonstrated that, as long as the resulting interfered signal is located in the constructive region of the modulation constellation, this increases the distance between the received signal and the detection thresholds, and an improved performance can be obtained. To illustrate this intuitively, in Fig. 2 we depict the constructive region for QPSK, where for simplicity and without loss of generality we focus on one quarter of the normalized QPSK constellation that corresponds to the constellation point $\left( {\frac{1}{{\sqrt 2 }} + \frac{1}{{\sqrt 2 }} \cdot j} \right)$. As can be observed, as long as the interfered signal ($\mathop {OB}\limits^ \to$ in Fig. 2) is located in the constructive region, the distance to the detection thresholds is increased, and an improved performance can be expected. The formulation of the optimization problem based on the constructive region will be introduced in the following.

\begin{figure}[!t]
\centering
\includegraphics[scale=0.3]{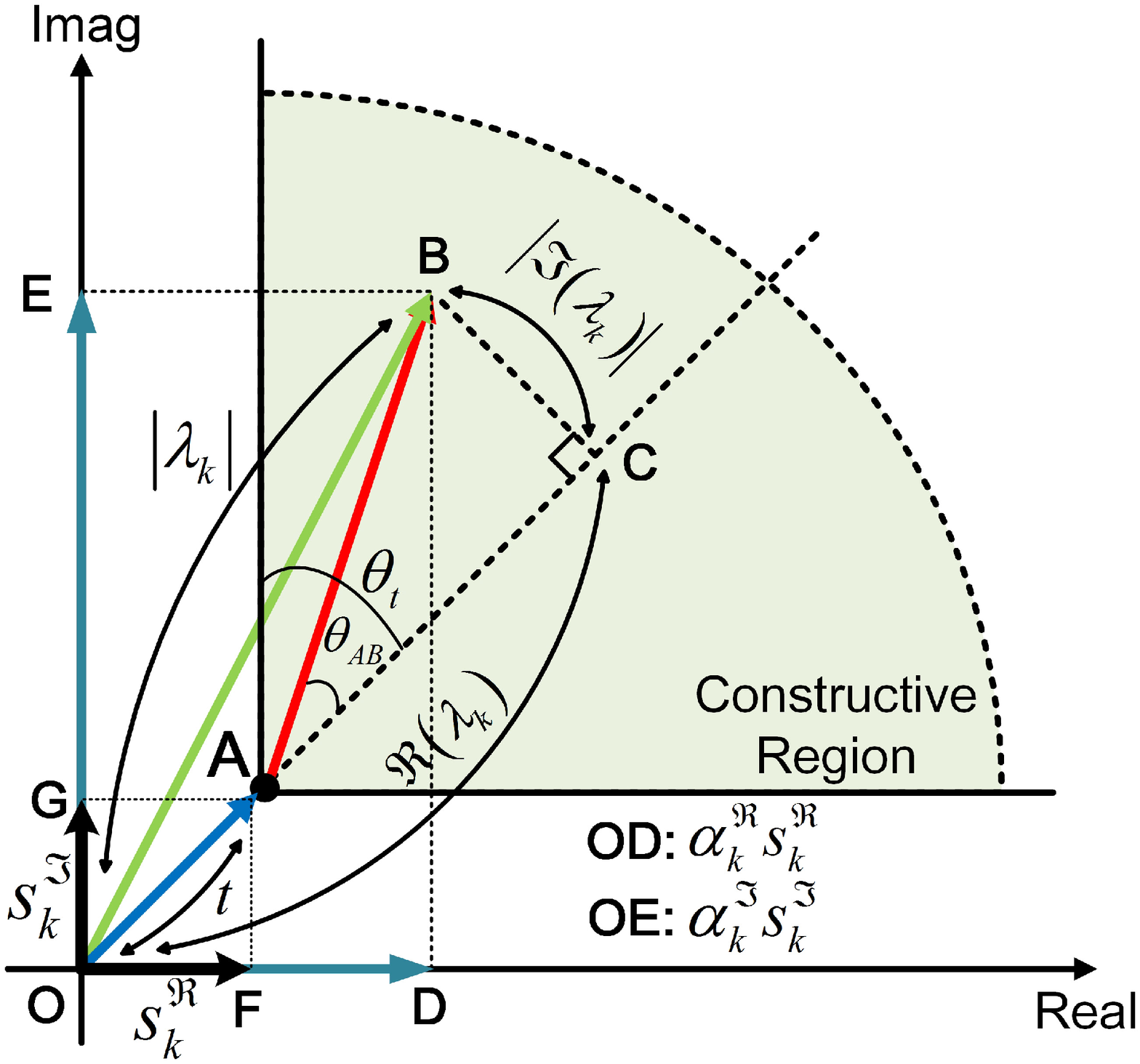}
\caption{Constructive interference and constructive region for QPSK}
\end{figure} 

\subsection{1-Bit Transmission Scheme - Linear Beamforming}
When a linear beamforming scheme is considered, the unquantized transmit signal vector can be expressed as
\begin{equation}
{{\bf{\hat x}}_T} = {\bf{Ws}}.
\end{equation}
To introduce the proposed scheme, we firstly decompose the channel matrix into 
\begin{equation}
{\bf{H}} = {\left[ {{\bf{h}}_1^T,{\bf{h}}_2^T,\cdots,{\bf{h}}_K^T} \right]^T},
\end{equation}
where each ${\bf h}_k \in {\cal C}^{1 \times N_t}$ denotes the channel vector of the $k$-th user. Then, the received signal for user $k$ can be obtained as
\begin{equation}
\begin{aligned}
{y_k} &= \sqrt P  \cdot {{\bf{h}}_k} {{{\bf{x}}_T}} + {n_k}\\
&= \sqrt P  \cdot {{\bf{h}}_k} {\cal Q}\left( {\bf Ws} \right) + {n_k},
\end{aligned}
\end{equation}
where $n_k$ is the $k$-th entry in $\bf n$. For the proposed quantized linear approach in this paper, the unquantized beamforming matrix $\bf W$ assuming infinite-precision DACs is firstly obtained, followed by the 1-bit quantization on the resulting transmit signal vector ${\bf \hat x}_T$.

%where $n_k$ is the $k$-th entry in $\bf n$. In this paper, we propose to maximize the distance between the constructive region and the detection thresholds such that the received signals are pushed as far as possible away from the detection thresholds. For the proposed quantized linear scheme, the unquantized beamforming matrix $\bf W$ assuming infinite-precision DACs is firstly obtained, followed by the quantization on the resulting transmit signal vector ${\bf \hat x}_T$ to satisfy the constraint on the output signals of 1-bit DACs. 

To formulate the desired optimization problem, let us firstly study the analytical constructive interference conditions. In Fig. 2, without loss of generality we denote $\mathop {OA}\limits^ \to = t \cdot {s_k}$ and $t = | \mathop {OA}\limits^ \to|$ is the objective to be maximized. We assume the node `B' denotes the noiseless received signal $\left( {{{\bf{h}}_k}{\bf{Ws}} }\right)$ that is located in the constructive region, and we further denote $\mathop {OB}\limits^ \to   = {\lambda _k}{s_k}$, where $\lambda_k$ is an introduced complex variable with $| \mathop {OB}\limits^ \to | = \left| {{\lambda _k}} \right|$. We can then obtain that
\begin{equation}
\mathop {OB}\limits^ \to   = {{\bf{h}}_k}{\bf{Ws}} = {\lambda _k}{s_k}.
\end{equation}
Based on the fact that $\mathop {OC}\limits^ \to$ and $\mathop {CB}\limits^ \to$ are perpendicular, we can further obtain $\mathop {OC}\limits^ \to$ and $\mathop {CB}\limits^ \to$, expressed as
\begin{equation}
\mathop {OC}\limits^ \to   = \Re \left( {{\lambda _k}} \right){s_k}, {\kern 3pt} \mathop {CB}\limits^ \to   = j \cdot \Im \left( {{\lambda _k}} \right){s_k},
\end{equation}
where geometrically the imaginary unit `$j$' denotes a phase rotation of ${90^{\rm{o}}}$ along the anti-clockwise direction. As the nodes `O', `A', and `C' are co-linear, we can then express $\mathop {AC}\limits^ \to$ as
\begin{equation}
\mathop {AC}\limits^ \to   = \left[ {\Re \left( {{\lambda _k}} \right) - t} \right]{s_k}.
\end{equation}
Based on the expression of $\mathop {AC}\limits^ \to$ and $\mathop {CB}\limits^ \to$, $\tan {\theta _{AB}}$ is obtained as
\begin{equation}
\tan {\theta _{AB}} = \frac{{\mathop {|CB|}\limits^ \to  }}{{\mathop {|AC|}\limits^ \to  }} = \frac{{\left| {\Im \left( {{\lambda _k}} \right){s_k}} \right|}}{{\left| {\left[ {\Re \left( {{\lambda _k}} \right) - t} \right]{s_k}} \right|}} = \frac{{\left| {\Im \left( {{\lambda _k}} \right)} \right|}}{{\Re \left( {{\lambda _k}} \right) - t}}.
\end{equation}
In Fig. 2, it is geometrically observed that to have node `B' located in the constructive region is equivalent to the following condition:
\begin{equation}
\begin{aligned}
&{\theta _{AB}} \le {\theta _t} \\
\Rightarrow & \tan {\theta _{AB}} \le \tan {\theta _t} \\
\Rightarrow & \frac{{\left| {\Im \left( {{\lambda _k}} \right)} \right|}}{{\Re \left( {{\lambda _k}} \right) - t}} \le \tan {\theta _t} \\
\Rightarrow & \left[ {\Re \left( {{\lambda _k}} \right) - t} \right]\tan {\theta _t} \ge \left| {\Im \left( {{\lambda _k}} \right)} \right|.
\end{aligned}
\end{equation}
For $\cal M$-PSK modulation, based on the geometry of the modulation constellation it is easy to obtain the threshold angle $\theta_t$, given by
\begin{equation}
{\theta _t} = \frac{\pi }{\cal M}.
\end{equation}
We can then formulate the optimization for the unquantized linear beamforming as
%With the obtained conditions to achieve constructive interference in (8) and (12), the unquantized linear beamforming optimization that maximizes the distance between the constructive region and the detection thresholds can be formulated as
\begin{equation}
\begin{aligned}
&\mathcal{P}_1: {\kern 3pt} \mathop {\max }\limits_{{{\bf{W}}}} {\kern 3pt} t \\
&{\kern 0pt} s. t. {\kern 10pt} {{\bf{h}}_k}{\bf Ws} = {\lambda _k}{s_k}, {\kern 3pt} \forall k \in {\cal K} \\
&{\kern 22pt} \left[ {\Re \left( {{\lambda _k}} \right) - t} \right]\tan {\theta _t} \ge \left| {\Im \left( {{\lambda _k}} \right)} \right|, {\kern 3pt} \forall k \in {\cal K}\\
&{\kern 22pt} \left\| {{\bf{Ws}}} \right\|_F \le \sqrt {p_0}\\
&{\kern 24pt} t \ge 0
\end{aligned}
\end{equation}
where ${\cal K}=\left\{ {1,2,\cdots,K} \right\}$, and $\left\| {{\bf{Ws}}} \right\|_F \le \sqrt {p_0}$ is the instantaneous power constraint on the beamformer as the beamforming is dependent on the data symbols. Due to the existence of the subsequent 1-bit quantization operation, $p_0$ in ${\cal P}_1$ can be any positive value, and this will not have an impact on the final obtained quantized signal vector ${\bf x}_T$. ${\cal P}_1$ is a second-order cone programming (SOCP) optimization, and we can further obtain the following proposition in the case of massive MIMO.

%${\cal P}_1$ is a second-order cone programming (SOCP) optimization problem and can be efficiently solved with convex optimization tools, where we can further obtain the following proposition in the case of massive MIMO.

$\bf Proposition$: In the case of massive MIMO, the optimality conditions for each $\lambda_k$ and $t$ of the optimization problem ${\cal P}_1$ are obtained as

\begin{enumerate}
\item $\Im \left( {{\lambda _k^*}} \right) = 0$, $\forall k \in {\cal K}$;
\item $t^* = {\lambda _1^*} = {\lambda _2^*} = \cdots = {\lambda _K^*} = \sqrt {\frac{{{N_t} \cdot {p_0}}}{K}}$.
\end{enumerate}

$\bf Proof$: We prove the above proposition by analyzing the optimization problem ${\cal P}_1$ with the Lagrangian approach. We firstly transform ${\cal P}_1$ into a standard minimization problem, given by 
\begin{equation}
\begin{aligned}
&\mathcal{P}_2: {\kern 3pt} \mathop {\min }\limits_{{{\bf{w}}_i}} {\kern 3pt} -t \\
&{\kern 0pt} s. t. {\kern 10pt} {{\bf{h}}_k}\sum\limits_{i = 1}^K {{{\bf{w}}_i}{s_i}}  - {\lambda _k}{s_k} = 0, {\kern 3pt} \forall k \in {\cal K} \\
&{\kern 22pt} \left| {\Im \left( {{\lambda _k}} \right)} \right| - \left[ {\Re \left( {{\lambda _k}} \right) - t} \right]\tan {\theta _t} \le 0, {\kern 3pt} \forall k \in {\cal K}\\
&{\kern 22pt} \sum\limits_{i = 1}^K {s_i^H{\bf{w}}_i^H{{\bf{w}}_i}{s_i}}  - {p_0} \le 0 
\end{aligned}
\end{equation}
where we note that the constraint on $t$ in ${\cal P}_1$ can be omitted in the above formulation, and we decompose ${\bf{W}} = \left[ {{{\bf{w}}_1},{{\bf{w}}_2},\cdots,{{\bf{w}}_K}} \right]$. We can then express the Lagrangian of ${\cal P}_2$ as \cite{r23}
\begin{equation}
\begin{aligned}
&{\cal L}\left( {{{\bf{w}}_i},t,{\delta _k},{\mu _k},{\mu _0}} \right) =  - t + \sum\limits_{k = 1}^K {{\delta _k}\left( {{{\bf{h}}_k}\sum\limits_{i = 1}^K {{{\bf{w}}_i}{s_i}}  - {\lambda _k}{s_k}} \right)}  \\
&+ {\mu _0}\left( {\sum\limits_{i = 1}^K {s_i^H{\bf{w}}_i^H{{\bf{w}}_i}{s_i}}  - {p_0}} \right) \\
&+ \sum\limits_{k = 1}^K {{\mu _k}\left[ {\left| {\Im \left( {{\lambda _k}} \right)} \right| - \Re \left( {{\lambda _k}} \right)\tan {\theta _t} + t \cdot \tan {\theta _t}} \right]},
\end{aligned}
\end{equation}
where ${\mu _0}$, ${\delta _k}$ and ${\mu _k}$ are the dual variables, and $\mu_0 \ge 0$, $\mu_k \ge 0$, $\forall k \in {\cal K}$. Based on the Lagrangian in (16), the KKT conditions for optimality are then obtained as
\begin{IEEEeqnarray}{rCl} 
\IEEEyesnumber
\frac{{\partial {\cal L}}}{{\partial t}} =  - 1 + \sum\limits_{k = 1}^K {{\mu _k}}  = 0 {\kern 30pt} \IEEEyessubnumber* \\
\frac{{\partial {\cal L}}}{{\partial {{\bf{w}}_i}}} = \left( {\sum\limits_{k = 1}^K {{\delta _k} \cdot {{\bf{h}}_k}} } \right){s_i} + {\mu _0} \cdot {\bf{w}}_i^H = {\bf{0}} {\kern 30pt} \\
{\mu _0}\left( {\sum\limits_{i = 1}^K {s_i^H{\bf{w}}_i^H{{\bf{w}}_i}{s_i}}  - {p_0}} \right) = 0 {\kern 30pt} \\
{\delta _k}\left( {{{\bf{h}}_k}\sum\limits_{i = 1}^K {{{\bf{w}}_i}{s_i}}  - {\lambda _k}{s_k}} \right) = 0, {\kern 2pt} \forall k \in {\cal K}{\kern 30pt} \\
{\mu _k}\left[ {\left| {\Im \left( {{\lambda _k}} \right)} \right| - \Re \left( {{\lambda _k}} \right)\tan {\theta _t} + t \cdot \tan {\theta _t}} \right] = 0, {\kern 2pt} \forall k \in {\cal K} {\kern 30pt}
\end{IEEEeqnarray}
Based on (17b), firstly it is easily obtained that $\mu_0 \ne 0$ which with the fact that $\mu_0 \ge 0$ further leads to $\mu_0 > 0$. Then, we can obtain ${\bf w}_i^H$ as
\begin{equation}
{\bf{w}}_i^H =  - \frac{1}{{{\mu _0}}} \cdot \left( {\sum\limits_{k = 1}^K {{\delta _k}{{\bf{h}}_k}} } \right){s_i}, {\kern 2pt} \forall i \in {\cal K}.
\end{equation}
By denoting 
\begin{equation}
{a_k} =  - \frac{{{\delta _k^H}}}{{{\mu _0}}}, {\kern 2pt} \forall k \in {\cal K},
\end{equation}
${\bf w}_i$ can be obtained from (18) and expressed as
\begin{equation}
{{\bf{w}}_i} = \left( {\sum\limits_{k = 1}^K {{a_k}{{\bf{h}}_k^H}} } \right){s_i^H}, {\kern 2pt} \forall i \in {\cal K}.
\end{equation}
Then, with the expression of each ${\bf w}_i$, the beamforming matrix $\bf W$ is obtained in a compact form as
\begin{equation}
\begin{aligned}
{\bf{W}} &= \left[ {{{\bf{w}}_1},{{\bf{w}}_2},\cdots,{{\bf{w}}_K}} \right] = \left( {\sum\limits_{k = 1}^K {{a_k}{\bf{h}}_k^H} } \right) \cdot \left[ {s_1^H,s_2^H,\cdots,s_K^H} \right] \\
&=\left[ {{\bf{h}}_1^H,{\bf{h}}_2^H,\cdots,{\bf{h}}_K^H} \right] {\left[ {{a_1},{a_2},\cdots,{a_K}} \right]^T} \left[ {s_1^H,s_2^H,\cdots,s_K^H} \right] \\
&={{\bf{H}}^H}{\bf{A}}{{\bf{s}}^H}.
\end{aligned}
\end{equation}
In order to obtain $\bf A$, we firstly rewrite (8) in a compact form, which is expressed as
\begin{equation}
{\bf{HWs}} = diag\left( {{\lambda _k}} \right){\bf{s}}.
\end{equation} 
Then, by substituting (21) into (22), the matrix $\bf A$ can be obtained based on $\lambda_k$, given by
\begin{equation}
\begin{aligned}
&{\bf{H}}{{\bf{H}}^H}{\bf{A}}{{\bf{s}}^H}{\bf{s}} = diag\left( {{\lambda _k}} \right){\bf{s}} \\
\Rightarrow & {\bf{A}} = \frac{1}{K} \cdot {\left( {{\bf{H}}{{\bf{H}}^H}} \right)^{ - 1}}diag\left( {{\lambda _k}} \right){\bf{s}}.
\end{aligned}
\end{equation}
The beamforming matrix $\bf W$ is then obtained as
\begin{equation}
{\bf{W}} = \frac{1}{K} \cdot {{\bf{H}}^H}{\left( {{\bf{H}}{{\bf{H}}^H}} \right)^{ - 1}}diag\left( {{\lambda _k}} \right){\bf{s}}{{\bf{s}}^H}.
\end{equation}
Based on the fact that $\mu_0 \ne 0$, it is obtained from (17c) that the power constraint of the optimization problem ${\cal P}_1$ is strictly active, which further leads to
\begin{equation}
\begin{aligned}
&{\left\| {{\bf{Ws}}} \right\|_F} = \sqrt {{p_0}}  \\
\Rightarrow {\kern 2pt} & tr\left\{ {{\bf{Ws}}{{\bf{s}}^H}{{\bf{W}}^H}} \right\} = {p_0} \\
\Rightarrow {\kern 2pt} & {{{\bf{s}}^H}{{\bf{W}}^H}{\bf{Ws}}} = {p_0}.
\end{aligned}
\end{equation}
Then, by substituting (24) into (25), we obtain that
\begin{equation}
\begin{aligned}
& {\kern 3pt} {{\bf{s}}^H}diag\left( {\lambda _k^H} \right){\left( {{\bf{H}}{{\bf{H}}^H}} \right)^{ - 1}}diag\left( {{\lambda _k}} \right){\bf{s}} = {p_0} \\
\Rightarrow & {\kern 2pt} {\rm{ve}}{{\rm{c}}^T}\left( {\lambda _k^H} \right)diag\left( {{{\bf{s}}^H}} \right){\left( {{\bf{H}}{{\bf{H}}^H}} \right)^{ - 1}}diag\left( {\bf{s}} \right){\rm{vec}}\left( {{\lambda _k}} \right) = {p_0} \\
\Rightarrow & \left[ {\lambda _1^H,\lambda _2^H,...,\lambda _K^H} \right] \cdot {\bf{T}} \cdot {\left[ {{\lambda _1},{\lambda _2},...,{\lambda _K}} \right]^T} = {p_0},
\end{aligned}
\end{equation}
where $\bf T$ is defined as
\begin{equation}
{\bf{T}} = diag\left( {{{\bf{s}}^H}} \right){\left( {{\bf{H}}{{\bf{H}}^H}} \right)^{ - 1}}diag\left( {\bf{s}} \right).
\end{equation}

In the case of massive MIMO, as ${N_t} \to \infty $, the favourable propagation property gives us that \cite{r10}
\begin{equation}
{\bf{H}}{{\bf{H}}^H} \approx {N_t} \cdot {\bf{I}} \Rightarrow {\left( {{\bf{H}}{{\bf{H}}^H}} \right)^{ - 1}} \approx \frac{1}{{{N_t}}} \cdot {\bf{I}},
\end{equation}
based on which $\bf T$ is further transformed into
\begin{equation}
{\bf{T}} \approx \frac{1}{{{N_t}}} \cdot diag\left( {{{\bf{s}}^H}} \right)diag\left( {\bf{s}} \right) = \frac{1}{{{N_t}}} \cdot {\bf{I}}.
\end{equation}
From the result in (29), (26) can be expanded and further transformed into
\begin{equation}
\frac{1}{{{N_t}}} \cdot \left( {{{\left| {{\lambda _1}} \right|}^2} + {{\left| {{\lambda _2}} \right|}^2} + \cdots + {{\left| {{\lambda _K}} \right|}^2}} \right) = {p_0}.
\end{equation}
To maximize $t$, as per (12) and (30) it is then easily obtained that the optimality is achieved when each $\lambda_k^*$ is real and identical, given by
\begin{equation}
{t^*} = \lambda _1^* = \cdots = \lambda _K^* = \sqrt {\frac{{{N_t} \cdot {p_0}}}{K}},
\end{equation}
which completes the proof. ${\kern 130pt} \blacksquare$

By substituting (31) into (24), the optimal beamforming matrix ${\bf W}^*$ can be expressed as
\begin{equation}
{{\bf{W}}^*} = {{\sqrt {\frac{{{N_t} \cdot {p_0}}}{K^3}} }} \cdot {{\bf{H}}^H}{\left( {{\bf{H}}{{\bf{H}}^H}} \right)^{ - 1}}{{\bf{s}}}{\bf{s}}^H.
\end{equation}
Then, with ${\bf W}^*$ obtained, the output signal vector that satisfies 1-bit DAC transmission is given as
\begin{equation}
\begin{aligned}
{\bf x}_T &= {\cal Q} \left( {{\bf W}^*{\bf s}} \right) \\
&= {\cal Q}\left( {\sqrt {\frac{{{N_t} \cdot {p_0}}}{{{K^3}}}} \cdot {{\bf{H}}^H}{{\left( {{\bf{H}}{{\bf{H}}^H}} \right)}^{ - 1}}{\bf{s}}{{\bf{s}}^H}{\bf{s}}} \right)\\
&={\cal Q}\left( {\sqrt {\frac{{{N_t} \cdot {p_0}}}{{{K}}}} \cdot {{\bf{H}}^H}{{\left( {{\bf{H}}{{\bf{H}}^H}} \right)}^{ - 1}}{\bf{s}}} \right).
\end{aligned}
\end{equation}
The intuition from the above proposition and (33) is that the quantized linear scheme based on the constructive interference is equivalent to the conventional quantized ZF scheme in the case of massive MIMO with 1-bit quantization, which suffers an error floor at high SNR \cite{r17}. This then motivates the proposed non-linear mapping scheme that achieves an improved performance in the following. 

\subsection{1-Bit Transmission Scheme - Non-linear Mapping}
%We proceed to introduce the optimization-based non-linear mapping scheme for massive MIMO with 1-bit DACs. The non-linear mapping scheme refers to the approach where, instead of designing a linear beamforming matrix that is multiplied to $\bf s$ before quantization, we directly design the quantized transmit signal vector ${\bf x}_T$. This approach was first described in \cite{r37}, and based on the constructive interference formulation in \cite{r33}. We employ this approach, to further design our low-complexity techniques in Section IV. The resulting optimization based on the constructive interference can be formulated as
We proceed to introduce the optimization-based non-linear mapping scheme for massive MIMO with 1-bit DACs. This approach was first described in \cite{r37}, and based on the constructive interference formulation in \cite{r33}. We employ this approach, to further design our low-complexity techniques in Section IV. The resulting optimization based on the constructive interference can be formulated as
\begin{equation}
\begin{aligned}
&\mathcal{P}_3: {\kern 3pt} \mathop {\max }\limits_{{{\bf{x}}_{T}}} {\kern 3pt} t \\
&{\kern 0pt} s. t. {\kern 10pt} {{\bf{h}}_k}{{\bf{x}}_{T}} = {\lambda _k}{s_k}, {\kern 3pt} \forall k \in {\cal K}\\
&{\kern 22pt} \left[ {\Re \left( {{\lambda _k}} \right) - t} \right]\tan {\theta _t} \ge \left| {\Im \left( {{\lambda _k}} \right)} \right|, {\kern 3pt} \forall k \in {\cal K}\\
&{\kern 24pt} {x_n} \in \left\{ { \pm \frac{1}{{\sqrt {2N_t} }} \pm \frac{1}{{\sqrt {2N_t} }}j} \right\}, {\kern 3pt} \forall n \in {\cal N}\\
&{\kern 24pt} t \ge 0
\end{aligned}
\end{equation}
It is observed that the optimization problem ${\cal P}_3$ is non-convex due to the output signal constraint for the 1-bit DACs in (34). To solve the above non-convex optimization, we adopt a two-step approach. 

\subsubsection{Relaxation}
In the first step, we relax the strict modulus constraint on each $x_n$ for both the real and imaginary part, and the resulting relaxed constraint can be expressed as
\begin{equation}
\left| {\Re \left( {{x_n}} \right)} \right| \le \frac{1}{{\sqrt {2{N_t}} }}, {\kern 3pt} \left| {\Im \left( {{x_n}} \right)} \right| \le \frac{1}{{\sqrt {2{N_t}} }}, {\kern 3pt} \forall n \in {\cal N}.
\end{equation}
The optimization problem ${\cal P}_3$ is then reformulated into a relaxed version ${\cal P}_4$, given by
\begin{equation}
\begin{aligned}
&\mathcal{P}_4: {\kern 3pt} \mathop {\max }\limits_{{{\bf \hat {x}}_{T}}} {\kern 3pt} t \\
&{\kern 0pt} s. t. {\kern 10pt} {{\bf{h}}_k}{{\bf \hat {x}}_{T}} = {\lambda _k}{s_k}, {\kern 3pt} \forall k \in {\cal K}\\
&{\kern 22pt} \left[ {\Re \left( {{\lambda _k}} \right) - t} \right]\tan {\theta _t} \ge \left| {\Im \left( {{\lambda _k}} \right)} \right|, {\kern 3pt} \forall k \in {\cal K}\\
&{\kern 22pt} \left| {\Re \left( {\hat x}_n \right)} \right| \le \frac{1}{{\sqrt {2N_t} }}, {\kern 3pt} \forall n \in {\cal N}\\
&{\kern 22pt} \left| {\Im \left( {\hat x}_n \right)} \right| \le \frac{1}{{\sqrt {2N_t} }}, {\kern 3pt} \forall n \in {\cal N}\\
&{\kern 24pt} t \ge 0
\end{aligned}
\end{equation}
where we denote $\hat x_n$ as the $n$-th entry in the relaxed transmit signal vector ${\bf \hat x}_T$. The resulting ${\cal P}_4$ is convex and can be solved with convex optimization tools.

\subsubsection{Normalization}
The solution obtained from the relaxed optimization ${\cal P}_4$ cannot always guarantee the equality on both the real and imaginary part of ${\hat x}_n$. To force the constraint of 1-bit transmission, the elements of the 1-bit DAC output ${\bf x}_T$ are obtained as
\begin{equation}
{x_n} = \frac{{\Re \left( {{{\hat x}_n}} \right)}}{{\sqrt {2{N_t}}  \cdot \left| {\Re \left( {{{\hat x}_n}} \right)} \right|}} + \frac{{\Im \left( {{{\hat x}_n}} \right)}}{{\sqrt {2{N_t}}  \cdot \left| {\Im \left( {{{\hat x}_n}} \right)} \right|}} \cdot j, {\kern 3pt} \forall n \in {\cal N}.
\end{equation}
We further note that, while we perform a relaxation on the 1-bit DAC constraint on each $x_n$ in ${\cal P}_3$, it turns out that most entries of the obtained ${\bf \hat x}_T$ from the relaxed problem ${\cal P}_4$ already meet the strict-equality requirement for 1-bit quantization, i.e. only a few entries of $\hat x_n$ need to be normalized. To evaluate the deviation of the relaxed optimization ${\cal P}_4$ from the original problem ${\cal P}_3$, we define $n_{\Re}$ and $n_{\Im}$ as the number of entries in the obtained ${\bf \hat x}_T$ whose absolute values are smaller than $\frac{1}{\sqrt {2N_t}}$ for the real and imaginary part, respectively. We further introduce
\begin{equation}
\eta  = \frac{{{n_\Re } + {n_\Im }}}{{2{N_t}}}
\end{equation}
as the ratio of the number of entries that do not satisfy the 1-bit transmission to the total number of entries in ${\bf \hat x}_T$, and this ratio therefore represents the deviation of the solution obtained by the relaxed problem from the original problem. We have $0 \le \eta  \le 1$, and ${\cal P}_4$ is equivalent to ${\cal P}_3$ if $\eta=0$. It is also observed that a smaller value of $\eta$ means that the relaxed optimization is closer to the original optimization. 

%Accordingly, there exist only a small number of ${\hat x}_n$ that need to be normalized to satisfy the constraint on the output signals for 1-bit DACs. Moreover, 

To study this numerically, we present the value of $\eta$ with respect to the number of antennas in Table I, where we have assumed a total number of $K=4$ users in the downlink system, and the result is based on 500 channel realizations. It is observed that the ratio $\eta$ decreases with the increase in the number of transmit antennas, which means that the solution obtained via the relaxed optimization problem ${\cal P}_4$ can be regarded as asymptotically optimal with an increasing number of transmit antennas in the case of massive MIMO.

\begin{table}[!t]
\begin{center}
\setlength{\arrayrulewidth}{0.25mm}
\renewcommand{\arraystretch}{1.25}
\scalebox{0.95}
{
\begin{tabular}{| c | c | c | c | c | }

    \hline
    Antenna number $N_t$ & 16 & 32 & 48 & 64 \\
    \hline
    Ratio $\eta$ & 20.52\% & 10.8\% & 7.28\% & 5.46\% \\
    \hline
    Antenna number $N_t$ & 80 & 96 & 112 & 128 \\
    \hline
    Ratio $\eta$ & 4.37\% & 3.65\% & 3.13\% & 2.73\% \\
    \hline
    
\end{tabular}
}
\end{center}
\label{tab:multicol}
\caption{$\eta$ with respect to the number of transmit antennas, $K=4$, 500 channel realizations}
\end{table}

\section{Proposed Low-Complexity Symbol Scaling Approach}
While the above non-linear mapping scheme can be relaxed into a convex optimization problem, the corresponding computational complexity is still prohibitively high as the variable dimension is equal to the number of transmit antennas. We study this mathematically and numerically in Section V and VI, respectively. Therefore in this section, we propose a three-stage symbol scaling scheme, which requires much reduced complexity for a comparable performance. It will be shown in the numerical results that for the small-scale MIMO systems, the low-complexity scheme even outperforms the optimization-based non-linear mapping scheme in Section III, since no relaxation or normalization is required for this scheme. %In the case of massive MIMO, it also achieves a comparable performance to the proposed non-linear mapping scheme.

\subsection{A New Look at the Constructive Interference Criteria}
To introduce the proposed symbol scaling scheme, we firstly perform a coordinate transformation on the formulation of the constructive interference constraint. To be specific, we firstly decompose each data symbol $s_k$ along its two corresponding detection thresholds of the modulation constellation, given by
\begin{equation}
{s_k} = \mathop {s_k^\Re }\limits^ \to + \mathop {s_k^\Im }\limits^ \to,
\end{equation}
where $\mathop {s_k^\Re }\limits^ \to$ and $\mathop {s_k^\Im }\limits^ \to$ are both complex values, and denoted as the two bases that are parallel to the two detection thresholds that correspond to the constellation point $s_k$. In the following, for simplicity we shall use $s_k^{\Re}$ and $s_k^{\Im}$ to denote the two bases. This is also shown geometrically in both Fig. 2 and Fig. 3 where we employ QPSK and 8-PSK modulation as examples, respectively. As observed in both figures, we decompose `OA' that represents the data symbol $s_k$ along its detection thresholds into `OF' and `OG'. For QPSK, based on Fig. 2 it is easy to observe that the real and imaginary axes are the detection thresholds, which leads to 
\begin{equation}
\mathop {OF}\limits^ \to = s_k^{\Re} = \frac{1}{\sqrt 2}, {\kern 3pt} \mathop {OG}\limits^ \to = s_k^{\Im} = \frac{1}{\sqrt 2} \cdot j
\end{equation}
for the corresponding constellation point `A'. For 8-PSK, `OD' and `OE' in Fig. 3 are the detection thresholds for the constellation point `A'. Then, with ${\theta _t} = {\pi  \mathord{\left/{\vphantom {\pi  8}} \right.\kern-\nulldelimiterspace} 8}$ for 8-PSK we can obtain the bases $s_k^{\Re}$ and $s_k^{\Im}$ that correspond to the constellation point `A' as
\begin{equation}
\begin{aligned}
\mathop {OF}\limits^ \to &= s_k^{{\mathop{\Re}\nolimits} } = \frac{{{e^{j \cdot \frac{\pi }{8}}}}}{{\left| {{{ {{e^{j \cdot \frac{\pi }{8}}}} }} + {{ {{e^{j \cdot \frac{{3\pi }}{8}}}} }}} \right|}} = {a_k} + {b_k} \cdot j, \\
\mathop {OG}\limits^ \to &= s_k^{{\mathop{\Im}\nolimits} } = \frac{{{e^{j \cdot \frac{{3\pi }}{8}}}}}{{\left| {{{{{e^{j \cdot \frac{\pi }{8}}}} }} + {{ {{e^{j \cdot \frac{{3\pi }}{8}}}} }}} \right|}} = {c_k} + {d_k} \cdot j.
\end{aligned}
\end{equation}
where $\left( {{a_k},{b_k}} \right)$ and $\left( {{c_k},{d_k}} \right)$ denote the coordinates of $s_k^{\Re}$ and $s_k^{\Im}$ in the conventional real-imaginary complex plane, respectively. The extension to other constellation points and higher order PSK modulations can be easily obtained in a similar way.

\begin{figure}[!t]
\centering
\includegraphics[scale=0.3]{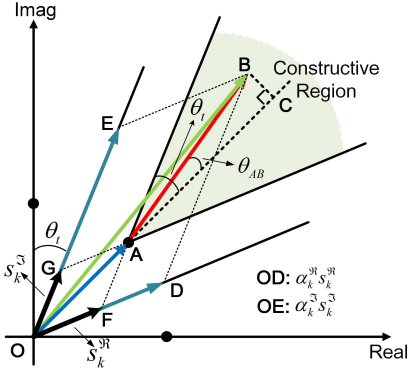}
\caption{Decomposition along the detection thresholds for 8-PSK}
\end{figure} 

Then for each $k$, instead of employing a complex scaling value $\lambda_k$ that is multiplied by $s_k$, with the above formulation (39)-(41) we introduce a symbol scaling approach where we decompose (8) along the two corresponding detection thresholds of $s_k$, given by
\begin{equation}
{{\bf{h}}_k}{{\bf{x}}_T} = \alpha _k^\Re s_k^\Re  + \alpha _k^\Im s_k^\Im,
\end{equation}
where 
\begin{equation}
\alpha _k^\Re \ge 0, {\kern 3pt} \alpha _k^\Im \ge 0, {\kern 3pt} \forall k \in {\cal K},
\end{equation}
are two introduced scaling factors that are multiplied to the bases $s_k^{\Re}$ and $s_k^{\Im}$, respectively. We can then observe that a larger value of $\alpha_k^{\Re}$ or $\alpha_k^{\Im}$ therefore represents a larger distance to the other detection threshold, and we further denote $\left( {\alpha _k^\Re ,\alpha _k^\Im } \right)$ as the coordinate of the node `B' in the complex plane expanded by the bases $s_k^{\Re}$ and $s_k^{\Im}$. By expanding (42) using the coordinate transformation, we can obtain the generic expression of $\alpha_k^{\Re}$ and $\alpha_k^{\Im}$ as a function of the transmit signal vector, given by (see Appendix)
\begin{equation}
\begin{aligned}
\alpha _k^\Re &= \frac{{{d_k}\Re \left( {{{\bf{h}}_k}} \right) - {c_k}\Im \left( {{{\bf{h}}_k}} \right)}}{{{a_k}{d_k} - {b_k}{c_k}}}{\bf{x}}_T^\Re  - \frac{{{d_k}\Im \left( {{{\bf{h}}_k}} \right) + {c_k}\Re \left( {{{\bf{h}}_k}} \right)}}{{{a_k}{d_k} - {b_k}{c_k}}}{\bf{x}}_T^\Im, \\
\alpha _k^\Im &=\frac{{{a_k}\Im \left( {{{\bf{h}}_k}} \right) - {b_k}\Re \left( {{{\bf{h}}_k}} \right)}}{{{a_k}{d_k} - {b_k}{c_k}}}{\bf{x}}_T^\Re  + \frac{{{a_k}\Re \left( {{{\bf{h}}_k}} \right) + {b_k}\Im \left( {{{\bf{h}}_k}} \right)}}{{{a_k}{d_k} - {b_k}{c_k}}}{\bf{x}}_T^\Im.
\end{aligned}
\end{equation}
In (44), for simplicity we have employed the following denotations
\begin{equation}
{\bf{x}}_T^\Re  = \Re \left( {{{\bf{x}}_T}} \right), {\kern 3pt} {\bf{x}}_T^\Im  = \Im \left( {{{\bf{x}}_T}} \right).
\end{equation}
By further denoting 
\begin{equation}
\begin{aligned}
&{{\bf{A}}_k} = \frac{{{d_k}\Re \left( {{{\bf{h}}_k}} \right) - {c_k}\Im \left( {{{\bf{h}}_k}} \right)}}{{{a_k}{d_k} - {b_k}{c_k}}}, {\kern 2pt} {{\bf{B}}_k} =  - \frac{{{d_k}\Im \left( {{{\bf{h}}_k}} \right) + {c_k}\Re \left( {{{\bf{h}}_k}} \right)}}{{{a_k}{d_k} - {b_k}{c_k}}}, \\
&{{\bf{C}}_k} = \frac{{{a_k}\Im \left( {{{\bf{h}}_k}} \right) - {b_k}\Re \left( {{{\bf{h}}_k}} \right)}}{{{a_k}{d_k} - {b_k}{c_k}}}, {\kern 2pt} {{\bf{D}}_k} = \frac{{{a_k}\Re \left( {{{\bf{h}}_k}} \right) + {b_k}\Im \left( {{{\bf{h}}_k}} \right)}}{{{a_k}{d_k} - {b_k}{c_k}}},
\end{aligned}
\end{equation}
the formulation of (44) is simplified into
\begin{equation}
\begin{aligned}
\alpha _k^\Re &= {{\bf{A}}_k}{\bf{x}}_T^\Re  + {{\bf{B}}_k}{\bf{x}}_T^\Im,\\
\alpha _k^\Im &= {{\bf{C}}_k}{\bf{x}}_T^\Re  + {{\bf{D}}_k}{\bf{x}}_T^\Im.
\end{aligned}
\end{equation}
By defining
\begin{equation}
{{\bf{R}}_k} = \left[ {\begin{array}{*{20}{c}}
{{{\bf{A}}_k}}&{{{\bf{B}}_k}}
\end{array}} \right], {\kern 3pt} {{\bf{I}}_k} = \left[ {\begin{array}{*{20}{c}}
{{{\bf{C}}_k}}&{{{\bf{D}}_k}}
\end{array}} \right],
\end{equation}
and 
\begin{equation}
{\bf{x}} = {\left[ {\begin{array}{*{20}{c}}
{{{\left( {{\bf{x}}_T^\Re } \right)}^T}}&{{{\left( {{\bf{x}}_T^\Im } \right)}^T}}
\end{array}} \right]^T}, {\kern 2pt} {\bf \Lambda}  = {\left[ {\alpha _1^\Re ,...,\alpha _K^\Re ,\alpha _1^\Im ,...,\alpha _K^\Im } \right]^T},
\end{equation}
(47) can be further expressed in a compact form as
\begin{equation}
{\bf \Lambda}  = {\bf{M}}{{\bf{x}}},
\end{equation}
where $\bf M$ is given by
\begin{equation}
{\bf{M}} = {\left[ {\begin{array}{*{20}{c}}
{{{\bf{R}}_1^T}}& \cdots &{{{\bf{R}}_K^T}}&{{{\bf{I}}_1^T}}& \cdots &{{{\bf{I}}_K^T}}
\end{array}} \right]^T}.
\end{equation}
With the above formulation, we can then construct the optimization problem as
\begin{equation}
\begin{aligned}
&\mathcal{P}_5: {\kern 3pt} \mathop {\max }\limits_{\bf{x}} \mathop {\min }\limits_l {\kern 3pt} \alpha _l \\
&{\kern 0pt} s. t. {\kern 10pt} {\bf \Lambda}  = {\bf{M}}{{\bf{x}}}\\
&{\kern 24pt} \alpha _l \ge 0, {\kern 3pt} \forall l \in {\cal L}\\
&{\kern 24pt} x_i^E \in \left\{ {\frac{1}{{\sqrt {2{N_t}} }}, - \frac{1}{{\sqrt {2{N_t}} }}} \right\}, {\kern 3pt} \forall i \in {\cal I}
\end{aligned}
\end{equation}
where we have omitted $\Re$ and $\Im$ in the expression of the entries of $\bf \Lambda$, and simply denote $\alpha_l$ as its $l$-th entry. In ${\cal P}_5$, ${\cal L} = \left\{ {1,2,\cdots,2K} \right\}$, $x_i^E$ denotes the $i$-th entry in ${\bf x}$ and ${\cal I} = \left\{ {1,2,\cdots,2{N_t}} \right\}$. The above optimization problem ${\cal P}_5$ is interpreted as follows: we aim to maximize the minimum value of $\alpha _l$ by selecting each $x_i^E$ as either $\frac{1}{{\sqrt {2{N_t}} }}$ or $-\frac{1}{{\sqrt {2{N_t}} }}$. With the above problem formulation, the relaxation-normalization process on the transmit signals is no longer needed. The above formulation motivates us to propose the following low-complexity scheme, which consists of three stages: an initialization stage, an allocation stage, and a refinement stage, all presented in the following in detail.

\subsection{Initialization Stage}
In the initialization stage, we directly select the value of $x_i^E$ for some $i$ by simple observation. To achieve this, we firstly decompose (50) into
\begin{equation}
{\bf \Lambda}  = \sum\limits_{i = 1}^{2{N_t}} {{{\bf{M}}_i}x_i^E}, \end{equation}
where we decompose $\bf M$ into
\begin{equation}
{\bf{M}} = \left[ {\begin{array}{*{20}{c}}
{{{\bf{M}}_1}}&{{{\bf{M}}_2}}& \cdots &{{{\bf{M}}_{2{N_t}}}}
\end{array}} \right],
\end{equation}
with each ${{\bf{M}}_i} \in {{\cal C}^{2K \times 1}}$. Then, we have the following observation.

$\bf Observation$: As long as all the entries of ${\bf M}_i$ share the same sign, then it is optimal to set the sign of the corresponding $x_i^E$ equal to that of ${{\bf{M}}_i}$, as in this case the values of each entry in $\bf \Lambda$ are guaranteed to increase. 

Then, the corresponding $x_i^E$ is obtained as
\begin{equation}
{{x}}_i^E = \frac{{\mathop{\rm sgn}} \left( {{{\bf{M}}_i}} \right)}{\sqrt {2N_t}}, {\kern 3pt} \forall i \in {\cal S},
\end{equation} 
where ${\mathop{\rm sgn}} \left(  {\bf a}  \right)$ defines a vector sign function and is only valid when each entry in the vector $\bf a$ has the same sign. $\cal S$ denotes the set that consists of the column indices of $\bf M$ that satisfy the sign-identity condition. We further introduce a column vector ${\bf{t}}$ that represents a temporary value of $\bf \Lambda$, given by
\begin{equation}
{\bf{t}} = \sum\limits_{i \in {\cal V}} {{{\bf{M}}_i}x_i^E},
\end{equation}
where the set $\cal V$ consists of the column indices of $\bf M$ whose corresponding $x_i^E$ have been allocated a value. We note that when $card\left( {\cal V} \right) = 2{N_t}$, we have ${\bf{t}} =\bf \Lambda$.

In the case that no column in $\bf M$ satisfies the sign-identity condition, in the initialization stage we select only one column, i.e. $card \left( {\cal S}\right)=1$, with the following criterion:
\begin{equation}
i = \mathop {\arg \max }\limits_{i \in {\cal I}} {\left\| {{{\bf{M}}_i}} \right\|_1},
\end{equation}
which selects the column that has the maximum effect on the value of $\bf \Lambda$. Then, the value of the corresponding $x_i^E$ is set as
\begin{equation}
x_i^E = \frac{{{\mathop{\rm sgn}} \left( {{\left\| {{{\bf{M}}_i}} \right\|_1}} \right)}}{{\sqrt {2{N_t}} }}.
\end{equation}
In the initialization stage, we have ${\cal V}={\cal S}$ or $card \left( {\cal V} \right)=1$. We summarize the algorithm for the initialization stage in Algorithm 1.

\begin{algorithm}
  \caption{Initialization Stage}
  \begin{algorithmic}
    \State ${\bf input:}$ ${\bf s}$, $\bf H$
    \State ${\bf output:}$ ${\bf t}$, $\cal V$
    \State Decompose each $s_k=s_k^{\Re}+s_k^{\Im}$ based on modulation type; 
    \State Obtain $\bf M$ based on (42)-(51);
    \State Find ${\bf M}_i$ that satisfies the sign-identity condition;
    \State Obtain $\cal S$;
    \If {${\cal S} \ne \emptyset $}
    \State ${{x}}_i^E = \frac{{\mathop{\rm sgn}} \left( {{{\bf{M}}_i}} \right)}{\sqrt {2N_t}}$, $\forall i \in {\cal S}$;
    \State ${\cal V}={\cal S}$;
    \Else
    \State Obtain $i$ based on (57), $x_i^E = \frac{{{\mathop{\rm sgn}} \left( {{\left\| {{{\bf{M}}_i}} \right\|_1}} \right)}}{{\sqrt {2{N_t}} }}$;
    \State ${\cal V}=\left\{ {i} \right\}$;
    \EndIf
    \State Calculate $\bf t$ based on (56).
   \end{algorithmic}
\end{algorithm}

\subsection{Allocation Stage}
At this stage we allocate the value of each $x_i^E$ for the residual $i$ that belongs to $\cal W$, where we define the set $\cal W$ as
\begin{equation}
{\cal W}= \left\{ {i {\kern 2pt} | {\kern 2pt} i \in {\cal I} {\kern 2pt} {\rm and} {\kern 2pt} i \notin {\cal V}} \right\}.
\end{equation}
$\cal W$ consists of those $x_i^E$ whose values have not been allocated in the initialization stage. In the following allocation stage, we consider both a `Sum-Max' and a `Max-Min' criteria for the allocation scheme.

\subsubsection{Sum-Max}
For the allocation scheme based on the `Sum-Max' criterion, instead of considering a max-min optimization as in ${\cal P}_5$, we consider a sum-max optimization where the objective function is constructed as
\begin{equation}
{\cal F}\left( {{{\bf{x}}}} \right) = {\rm{sum}}\left( {\bf \Lambda} \right),
\end{equation}
where ${\rm sum}\left( {\bf a} \right)$ returns the sum of the entries in a column vector $\bf a$. Then, based on (50) the objective can be further transformed into
\begin{equation}
{\cal F}\left( {{{\bf{x}}}} \right) = {\bf{m}}{{\bf{x}}}= \sum\limits_{i = 1}^{2{N_t}} {{\bf{m}}\left( i \right)x_i^E},
\end{equation}
where ${\bf{m}} \in {{\cal C}^{1 \times 2{N_t}}}$ is the sum of the entries in each row of M. Each ${\bf{m}}\left( i \right)$ denotes the $i$-th entry in $\bf m$, given by
\begin{equation}
{\bf{m}}\left( i \right) = \sum\limits_{l = 1}^{2K} {{{\bf{M}}_i}\left( l \right)}.
\end{equation}
It is then easy to observe that ${\cal F}\left( {{{\bf{x}}}} \right)$ is maximized when the sign of each $x_i^E$ is the same as that of ${\bf{m}}\left( i \right)$, and therefore the optimal $x_i^E$ for the `Sum-Max' criterion is given by
\begin{equation}
x_i^E = \frac{{{\mathop{\rm sgn}} \left[ {{\bf{m}}\left( i \right)} \right]}}{{\sqrt {2{N_t}} }}, {\kern 3pt} \forall i \in {\cal W}.
\end{equation}
While the above solution guarantees that the sum of $\alpha_l$ is maximized, it does not specifically consider each value of $\alpha_l$, which may lead to performance loss. Indeed, it is possible that the value of one $\alpha_l$ can be very small or even negative. This is the reason why the refinement in Section IV-D is further introduced. The algorithm for the allocation stage based on `Sum-Max' is summarized in Algorithm 2.

\begin{algorithm}
  \caption{Allocation Stage - `Sum-Max'}
  \begin{algorithmic}
    \State ${\bf input:}$ ${\cal V}$, $\bf M$ 
    \State ${\bf output:}$ ${\bf x}_{\rm sum-max}$
    \State Calculate $\cal W$ based on (59);
    \State Calculate $\bf m$ and each ${\bf m}\left( i \right)$ based on (61), (62);
    \State Allocate $x_i^E = \frac{{{\mathop{\rm sgn}} \left[ {{\bf{m}}\left( i \right)} \right]}}{{\sqrt {2{N_t}} }}$, $\forall i \in {\cal W}$;
    \State Obtain $\bf x$, denoted as ${\bf x}_{\rm sum-max}$.
   \end{algorithmic}
\end{algorithm}

\subsubsection{Max-Min}
For the `Max-Min' allocation criterion, in each step we aim to improve the minimum value in $\bf \Lambda$ as much as possible. Denoting $q$ as the row index of the minimum entry in $\bf t$ obtained in the initialization stage, we have
\begin{equation}
{\bf{t}}\left( q \right) = \min \left( {\bf{t}} \right),
\end{equation}
where $\min \left( {\bf{t}} \right)$ returns the minimum value in $\bf t$. Subsequently, we iteratively select ${\bf M}_i$ with the largest absolute value in the $q$-th row, given by
\begin{equation}
\mathop {i = \arg \max }\limits_{i \in {\cal W}} \left| {{{\bf{M}}_i}\left( q \right)} \right|,
\end{equation}
and the corresponding $x_i^E$ is then obtained as
\begin{equation}
x_i^E = \frac{{{\mathop{\rm sgn}} \left[ {{{\bf{M}}_i}\left( q \right)} \right]}}{{\sqrt {2{N_t}} }}.
\end{equation}
Then, we update $\cal V$ and $\bf t$, and based on the updated $\bf t$ we repeat the above procedure until ${\cal V}={\cal I}$. This means that each entry in $\bf x$ has been allocated, and the algorithm for the allocation stage based on `Max-Min' is summarized in Algorithm 3.

\begin{algorithm}
  \caption{Allocation Stage - `Max-Min'}
  \begin{algorithmic}
    \State ${\bf input:}$ ${\cal V}$, $\bf M$, $\bf t$ 
    \State ${\bf output:}$ ${\bf x}_{\rm max-min}$
    \While {${\cal V} \ne {\cal I}$}
    \State Calculate $\cal W$ based on (59);
    \State Obtain $q$ that satisfies ${\bf{t}}\left( q \right) = \min \left( {\bf{t}} \right)$;
    \State Find $\mathop {i = \arg \max }\limits_{i \in {\cal W}} \left| {{{\bf{M}}_i}\left( q \right)} \right|$;
    \State Allocate $x_i^E = \frac{{{\mathop{\rm sgn}} \left[ {{{\bf{M}}_i}\left( q \right)} \right]}}{{\sqrt {2{N_t}} }}$;
    \State Update ${\cal V}$ and $\bf t$;
    \EndWhile
    \State Obtain $\bf x$, denoted as ${\bf x}_{\rm max-min}$.
   \end{algorithmic}
\end{algorithm}

\subsection{Refinement Stage}
In the refinement stage, we check whether the performance based on the obtained signal vector in the allocation stage can be further improved based on a greedy algorithm. To introduce the refinement process, we denote the obtained expanded 1-bit signal vector after the allocation stage as ${\bf x}$ (obtained based on either the `Sum-Max' or the `Max-Min' criterion). First, we sequentially change the sign of one entry (for example $x_i^E$) in ${\bf x}$ at a time while fixing the signs of other entries in $\bf x$, and denote the modified signal vector as ${\bf x}_{(i)}$. We then compare the minimum value in $\bf \Lambda$ obtained by the modified ${\bf x}_{(i)}$ with the minimum value in the original $\bf \Lambda$ obtained by ${\bf x}_{(0)}$. The sign of $x_i^E$ is selected as the one that returns a larger minimum value in $\bf \Lambda$. The refinement process is sequentially performed for each entry in ${\bf x}_{(0)}$. The algorithm for the refinement stage is then shown in Algorithm 4.

\begin{algorithm}
  \caption{Refinement Stage}
  \begin{algorithmic}
    \State ${\bf input:}$ ${\bf x}_{\rm sum-max}$ (or ${\bf x}_{\rm max-min}$)
    \State ${\bf output:}$ ${\bf x}_T$
    \State Denote ${\bf x}_{(0)}={\bf x}_{\rm sum-max}$ (or ${\bf x}_{\rm max-min}$);
    \For {$i = 1:2{N_t}$}
    \State Calculate ${\bf \Lambda}_{\left( 0 \right)}  = {\bf{M}}{{\bf{x}}_{\left( 0 \right)}}$;
    \State Obtain ${{\bf{x}}_{\left( i \right)}} = {\left[ {{x_1^E},...,{x_{i - 1}^E}, - {x_i^E},{x_{i + 1}^E},...,{x_{2N_t}^E}} \right]^T}$;
    \State Calculate ${\bf \Lambda}_{\left( i \right)}  = {\bf{M}}{{\bf{x}}_{\left( i \right)}}$;
    \If {$\min \left( {{{\bf \Lambda} _{\left( i \right)}}} \right) > \min \left( {{{\bf \Lambda} _{\left( 0 \right)}}} \right)$}
    \State ${x_i^E} \leftarrow  - {x_i^E}$;
    \State Update ${\bf x}_{(0)}$;
    \EndIf
    \EndFor
    \State Obtain ${\bf x}_T$ based on the updated ${\bf x}_{(0)}$.
   \end{algorithmic}
\end{algorithm}

The refinement stage is performed for the signal vectors obtained by both the `Sum-Max' and `Max-Min' criteria independently. The final output signal vector of the proposed symbol scaling scheme that generates the best performance is then selected between the signal vectors obtained with these two criteria. 

\subsection{Algorithm}
Based on the above description, the algorithm for the three-stage symbol scaling scheme is summarized in Algorithm 5, where the final output signal vector of the proposed symbol scaling scheme that generates the best performance is selected within the signal vectors obtained by the `Sum-Max' and `Max-Min' criteria.

\begin{algorithm}
  \caption{The Proposed Symbol Scaling Scheme}
  \begin{algorithmic}
    \State ${\bf input:}$ ${\bf s}$, $\bf H$
    \State ${\bf output:}$ ${\bf x}_T$
    
    \State $\bf Initialization$ $\bf Stage$
    \State Obtain ${\cal V}$, $\bf M$, and $\bf t$ with Algorithm 1;
    
    \State $\bf Allocation$ $\bf Stage$
    \State $\bf 1. `Sum-Max' :$
    \State Obtain ${\bf x}_{\rm sum-max}$ with Algorithm 2; 
    
    \State $\bf 2. `Max-Min' :$
    \State Obtain ${\bf x}_{\rm max-min}$ with Algorithm 3;
    
    \State $\bf Refinement$ $\bf Stage$ 
    \State Update both ${\bf x}_{\rm sum-max}$ and ${\bf x}_{\rm max-min}$ with Algorithm 4;
    \State Calculate ${{\bf \Lambda} _{\rm{s}}} = {\bf{M}}{{\bf{x}}_{{\rm{sum - max}}}}$ and ${{\bf \Lambda} _{\rm{m}}} = {\bf{M}}{{\bf{x}}_{{\rm{max - min}}}}$;
    \If {$\min \left( {{{\bf \Lambda} _{\rm{s}}}} \right) > \min \left( {{{\bf \Lambda} _{\rm{m}}}} \right)$}
    \State ${\bf x}={\bf x}_{\rm sum-max}$;
    \Else
    \State ${\bf x}={\bf x}_{\rm max-min}$;
    \EndIf
    
    \State Decompose ${\bf{x}} = {\left[ {\begin{array}{*{20}{c}}
{{{\left( {{\bf{x}}_T^\Re } \right)}^T}}&{{{\left( {{\bf{x}}_T^\Im } \right)}^T}}
\end{array}} \right]^T}$; 
    \State Output ${{\bf{x}}_T} = {\bf{x}}_T^\Re  + {\bf{x}}_T^\Im  \cdot j$.
   \end{algorithmic}
\end{algorithm}

\begin{table*}[!t]
\begin{center}
\setlength{\arrayrulewidth}{0.25mm}
\renewcommand{\arraystretch}{1.35}
\scalebox{0.95}
{
\begin{tabular}{|c|c|c|c|c|c|}
   \hline
   \multirow{2}{*}{Antenna Number} & \multicolumn{4}{|c|}{Schemes} \\ \cline{2-5}
   & Exhaustive Search & Proposed Non-linear Mapping ${\cal P}_4$ & Proposed Symbol Scaling & Non-linear Pokemon, $n_{\max}=20$ \\    
   \hline
   64 & ${\cal O}\left\{ {1.39 \times {10^{42}}} \right\}$ & ${\cal O}\left\{ {2.83 \times {10^{7}}} \right\}$ & ${\cal O}\left\{ {7.9 \times {10^{4}}} \right\}$ & ${\cal O}\left\{ {6.6 \times 10^{5}} \right\}$ \\
   \hline
   96 & ${\cal O}\left\{ {3.86 \times {10^{61}}} \right\}$ & ${\cal O}\left\{ {1.11 \times 10^{8}} \right\}$ & ${\cal O}\left\{ {1.74 \times {10^{5}}} \right\}$ & ${\cal O}\left\{ {1.48 \times 10^{6}} \right\}$ \\
   \hline   
   128 & ${\cal O}\left\{ {9.49 \times {10^{80}}} \right\}$ & ${\cal O}\left\{ {2.94 \times 10^{8}} \right\}$ & ${\cal O}\left\{ {3.05 \times 10^{5}} \right\}$ & ${\cal O}\left\{ {2.63 \times 10^{6}} \right\}$ \\
   \hline
   256 & ${\cal O}\left\{ {2.20 \times {10^{158}}} \right\}$ & ${\cal O}\left\{ {3.18 \times 10^{9}} \right\}$ & ${\cal O}\left\{ {1.2 \times 10^{6}} \right\}$ & ${\cal O}\left\{ {1.05 \times 10^{7}} \right\}$ \\
   \hline
\end{tabular}
}
\end{center}
\label{tab:multicol}
\caption{Comparison of the computational costs of different schemes, $K=8$}
\end{table*}

\section{Computational Complexity Analysis}
In this section we study the computational costs of the proposed schemes in terms of the floating-point operations required. As a reference, we also study the complexity of the exhaustive search scheme and the non-linear `Pokemon' scheme in \cite{r20}. The computational costs of all considered approaches are calculated based on real multiplications and additions.

\subsection{Exhaustive Search}
For massive MIMO transmission with 1-bit quantization, the output signal on each antenna element has 4 potential values, and for each signal combination it takes $4KN_t$ multiplications and $4KN_t$ additions to compute $\bf \Lambda$ based on (50) as ${\bf M} \in {\cal C}^{2K\times2N_t}$. Therefore, the complexity of the exhaustive search scheme is obtained as
\begin{equation}
{{\rm{C}}_{\rm E}} = {\cal O}\left\{ {{8KN_t\cdot4^{{N_t}}}} \right\} = {\cal O}\left\{ {{8KN_t\cdot2^{2{N_t}}}} \right\}.
\end{equation}
It is easy to conclude that in the case of massive MIMO, the exhaustive search scheme is inapplicable due to the overwhelmingly high computational cost.

\subsection{Optimization-based Non-linear Mapping ${\cal P}_4$}
For the proposed non-linear mapping scheme, in the relaxation stage the complexity is dominated from solving the relaxed convex problem ${\cal P}_5$ via the interior-point method \cite{r23}. It has been shown in \cite{r24} that the arithmetic complexity of the interior-point method is given by
\begin{equation}
{{\rm{C}}_{\rm I}} = {\cal O}\left\{ {{{\left( {M + N} \right)}^{1.5}}{M^2}} \right\},
\end{equation}
where $M$ is the dimension of the variable, and $N$ is the number of constraints. Based on the real representation ${\cal P}_5$, we obtain $M=2N_t$ and $N=2K$, which leads to
\begin{equation}
\begin{aligned}
{\rm C}_{\rm N}^1 &= {\cal O}\left\{ {{{(2K + 2{N_t})}^{1.5}}{{\left( {2{N_t}} \right)}^2}} \right\}\\
& = {\cal O} \left\{ {8\sqrt 2 {{\left( {K + {N_t}} \right)}^{1.5}}N_t^2} \right\}.
\end{aligned}
\end{equation}
In the normalization stage, the dominant complexity comes from the search for the signals that do not satisfy the output constraint for the 1-bit transmission. There are a total number of $2N_t$ entries in ${\bf \hat x}_T$ including both the real and imaginary part, and therefore a one-dimensional search of $2N_t$ entries is required. Then, the resulting complexity is obtained as
\begin{equation}
{\rm{C}}_{\rm{N}}^2 = O\left\{ {2{N_t}} \right\},
\end{equation}
which leads to the total computational cost for the optimization-based non-linear mapping scheme as
\begin{equation}
{{\rm{C}}_{\rm{N}}}  = {\rm{C}}_{\rm{N}}^1 + {\rm{C}}_{\rm{N}}^2 = {\cal O} \left\{ {8\sqrt 2 {{\left( {K + {N_t}} \right)}^{1.5}}N_t^2} \right\} + {\cal O}\left\{ {2{N_t}} \right\}.
\end{equation}
In the case of massive MIMO where $N_t$ is large, we have the following approximation:
\begin{equation}
{{\rm{C}}_{\rm{N}}}\approx {\cal O} \left\{ {8\sqrt 2 {{\left( {K + {N_t}} \right)}^{1.5}}N_t^2} \right\}.
\end{equation}

\subsection{Symbol Scaling Scheme}
For the proposed symbol scaling approach, in the following we calculate its computational cost for each stage. For both allocation criteria, the main computational cost in the initialization and allocation stage comes from the calculation of ${\bf t}\in {\cal C}^{2N_t\times1}$ based on (56). While the calculation of $\bf t$ is not necessary for the `Sum-Max' criterion, we note that $\bf t$ is required in the refinement stage. Each additional $\left( {{{\bf{M}}_i}x_i^E} \right)$ term that is added to $\bf t$ requires $2N_t$ multiplications and $2N_t$ additions, and $\bf t$ is updated $2K$ times after the allocation stage, where we note ${\bf M} \in {\cal C}^{2K \times 2N_t}$. The resulting computation cost is
\begin{equation}
{\rm {C}}_{\rm L}^1 = {\cal O}\left\{ {2K\left( {2{N_t} + 2{N_t}} \right)} \right\} ={\cal O} \left\{ {8K{N_t}} \right\}.
\end{equation}
Moreover, for the `Max-Min' allocation criterion, we need to iteratively allocate the value for the residual $x_i^E$, which introduces an additional computational cost for `Max-Min' in the allocation stage. Since $card\left({\cal V}\right)$ is difficult to obtain analytically in the initialization stage, we consider a worst-case complexity where $card\left({\cal V}\right)=1$, and in each iteration obtaining $q$ and $i$ in Algorithm 3 requires $2K$ and $2N_t$ operations, respectively. The required number of computations is thus
\begin{equation}
{\rm {C}}_{\rm L}^2={\cal O}\left\{ {\left( {2{N_t} - 1} \right)\left( {2K + {2N_t}} \right)} \right\} \approx {\cal O}\left\{ {4N_t^2 + 4K{N_t}} \right\}
\end{equation}
in the case of massive MIMO. In the refinement stage, it is easy to observe that the initial ${\bf \Lambda}_{\left( {0}\right)} = {\bf t}$. Then, in each iteration of Algorithm 4 we only need to calculate the corresponding ${{\bf{M}}_i} \cdot \left( {-x_i^E} \right)$ and include it in ${\bf \Lambda}_{\left( {i}\right)}$. For each $x_i^E$ this takes $2N_t$ multiplications and $2N_t$ additions, and therefore the computational cost for the refinement stage is 
\begin{equation}
{\rm {C}}_{\rm L}^3 = {\cal O}\left\{ {2{N_t}\left( {2{N_t} + 2{N_t}} \right)} \right\} = {\cal O} \left\{ {8N_t^2} \right\}.
\end{equation}
Based on Algorithm 5, both ${\bf x}_{\rm sum-max}$ and ${\bf x}_{\rm max-min}$ should be refined. Accordingly, we can obtain the total computational cost for the proposed symbol scaling approach as
\begin{equation}
\begin{aligned}
{{\rm{C}}_{\rm{L}}} &= {\rm{C}}_{\rm{L}}^1 + {\rm{C}}_{\rm{L}}^2+ 2 {\rm{C}}_{\rm{L}}^3 \\
&= {\cal O} \left\{ {8K{N_t}} \right\} + {\cal O}\left\{ {4N_t^2 + 4K{N_t}} \right\} + {\cal O} \left\{ {2\left( {8N_t^2} \right)} \right\} \\
&= {\cal O} \left\{ {20N_t^2+ 12K{N_t}} \right\} .
\end{aligned}
\end{equation}

\subsection{Pokemon}
As a comparison, we also include the complexity of the non-linear `Pokemon' scheme proposed in \cite{r20}. The `Pokemon' approach is based on biconvex relaxation, whose performance is dependent on the number of required iterations. Based on \cite{r20}, in each iteration we need to first calculate a vector ${\bf q} \in {\cal C}^{2N_t \times 1}$ based on ${\bf{q}} = {\bf{Ux}}$ where ${\bf U} \in {\cal C}^{2N_t\times2N_t}$, and then update the signal vector ${\bf x} \in {\cal C}^{2N_t\times1}$ with a projection function. The calculation of $\bf q$ requires a total of $4N_t^2$ multiplications and $4N_t^2$ additions, while the update of $\bf x$ requires $4N_t$ multiplications. Assuming a maximum number of iterations $n_{\rm max}$, this leads to the total computational cost for `Pokemon' as
\begin{equation}
\begin{aligned}
{{\rm{C}}_{\rm{P}}} &= {\cal O} \left\{ {{n_{\max }}\left( {4N_t^2 + 4N_t^2 + 4{N_t}} \right)} \right\} \\
&={\cal O} \left\{ {n_{\rm max} \left( {8N_t^2 + 4{N_t}}\right)} \right\}.
\end{aligned}
\end{equation}
Comparing the computational cost of `Pokemon' with the proposed symbol scaling method, we have
\begin{equation}
\frac{{{{\rm{C}}_{\rm{L}}}}}{{{{\rm{C}}_{\rm{P}}}}} = \frac{{{\cal O}\left\{ {20N_t^2 + 12K{N_t}} \right\}}}{{{\cal O}\left\{ {n_{\rm max} \left( {8N_t^2 + 4{N_t}} \right)} \right\}}} = {\cal O} \left\{ {\frac{{5{N_t} + 3K}}{{{n_{\max }}\left( {2{N_t} + 1} \right)}}} \right\}.
\end{equation}
In the case of massive MIMO where $K$ is finite while the antenna number $N_t \rightarrow \infty$, (78) is further transformed into
\begin{equation}
\frac{{{{\rm{C}}_{\rm{L}}}}}{{{{\rm{C}}_{\rm{P}}}}} = {\cal O}\left\{ {\frac{{5 + \frac{{3K}}{{{N_t}}}}}{{{n_{\max }}\left( {2 + \frac{1}{{{N_t}}}} \right)}}} \right\} \approx {\cal O}\left\{ {\frac{2.5}{{{n_{\max }}}}} \right\}.
\end{equation}

To numerically study the complexity gains of the proposed symbol scaling method, in Table II we show the number of floating-point operations required as the number of transmit antennas increases, where for `Pokemon' we employ $n_{\max}=20$ following \cite{r20}. As can be seen, the computational cost of the proposed non-linear mapping scheme is higher than that of the proposed symbol scaling approach and the `Pokemon' method, while the number of operations required for the proposed symbol scaling approach is approximately 12$\%$ of the number of operations for `Pokemon'. 

\section{Numerical Results}
In this section we present the numerical results of the proposed approaches based on Monte Carlo simulations. In each plot, the transmit SNR is defined as $\rho  = {P \mathord{\left/ {\vphantom {P {{\sigma ^2}}}} \right. \kern-\nulldelimiterspace} {{\sigma ^2}}}$. Both QPSK and 8-PSK modulations are considered in the numerical results. We compare our proposed methods with both the quantized linear approaches and the non-linear mapping algorithms, and for clarity the following abbreviations are used throughout this section:

\begin{enumerate}
\item `ZF-FD': Unquantized ZF beamforming with infinite-precision DACs;
\item `ZF 1-Bit': Quantized ZF approach with 1-bit DACs introduced in \cite{r17};
\item `MMSE': MMSE-based quantized linear scheme in \cite{r18};
\item `Pokemon, $n_{\max}=K$': Non-linear Pokemon algorithm proposed in \cite{r20} with $K$ iterations;
\item `Constructive': Proposed non-linear mapping scheme ${\cal P}_4$ in Section III-B;
\item `sum-max': Proposed symbol scaling approach based on the `sum-max' allocation scheme with Algorithm 1, 2 and 4;
\item `max-min': Proposed symbol scaling approach based on the `max-min' allocation scheme with Algorithm 1, 3 and 4;
\item `Symbol Scaling': Proposed symbol scaling method obtained via Algorithm 5 where we select the best signal vector out of `sum-max' or `max-min' criteria.
\end{enumerate}

\begin{figure}[!t]
\centering
\includegraphics[scale=0.45]{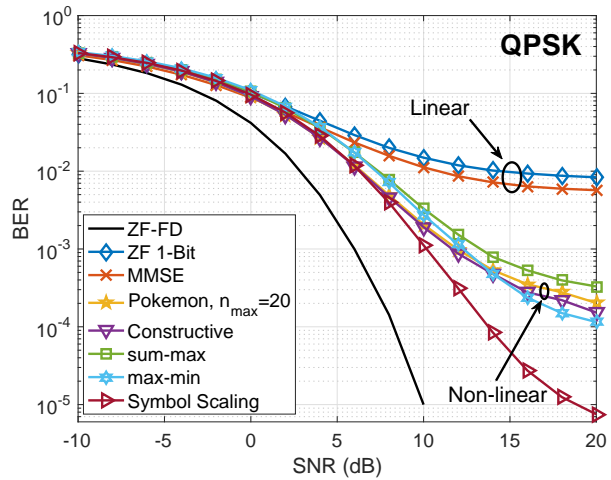}
\caption{BER v.s. transmit SNR, $N_t=8$, $K=2$, $n_{\max}=20$, QPSK}
\end{figure} 

In Fig. 4, we firstly consider a moderate scale MIMO with a total number of $N_t=8$ transmit antennas at the BS and $K=2$ single-antenna users in the system. For approaches with 1-bit quantization, we observe that the proposed symbol scaling scheme based on Algorithm 5 achieves the best BER performance, while both the proposed non-linear mapping scheme and `Pokemon' achieve an inferior performance. This is because both the non-linear mapping method and the `Pokemon' approach involve the relaxation-normalization process. For small-scale MIMO systems, based on Table I we can infer that $\eta$ will be large in this case, which means that the deviation of the solution obtained by the relaxation-normalization process from the solution of the original 1-bit optimization problem is large, and the normalization process may lead to further detection errors. For the proposed symbol scaling scheme, the performance is promising since we directly select the quantized signal for each antenna element and therefore no relaxation or quantization is needed.

\begin{figure}[!t]
\centering
\includegraphics[scale=0.45]{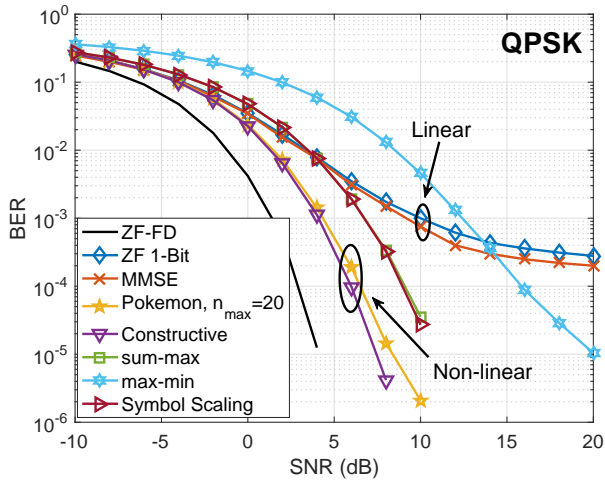}
\caption{BER v.s. transmit SNR, $N_t=128$, $K=16$, $n_{\max}=20$, QPSK}
\end{figure} 

We then consider a massive MIMO system with $N_t=128$ transmit antennas and $K=16$ users in Fig. 5. In the case of massive MIMO, all the schemes can achieve a lower BER thanks to the large number of antennas at the BS, and generally non-linear schemes outperform linear schemes. For approaches with 1-bit DACs, the proposed non-linear mapping method outperforms the non-linear `Pokemon' algorithm and achieves the best BER performance. As for the proposed low-complexity symbol scaling scheme, by comparing Fig. 4 and Fig. 5, we can observe that the `Max-Min' criterion is most suitable for small-scale MIMO systems, while the `Sum-Max' criterion is more favourable for massive MIMO systems. Moreover, while we have observed around a 2dB SNR loss compared to the `Pokemon' algorithm in the case of massive MIMO, its computational cost is approximately 12$\%$ of that for Pokemon in this scenario, which is shown mathematically in Table II and will be shown numerically in Fig. 7.

\begin{figure}[!t]
\centering
\includegraphics[scale=0.45]{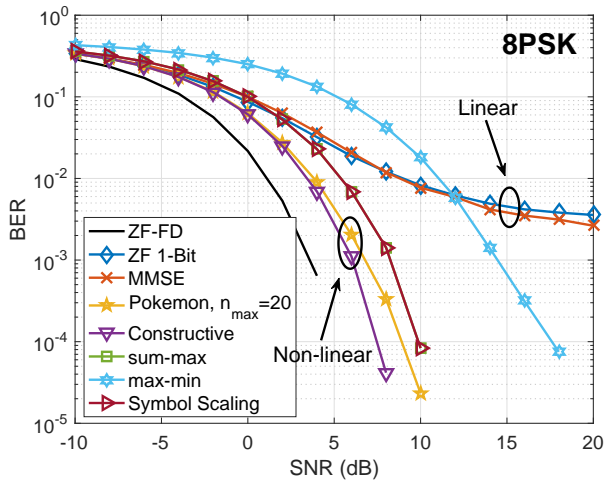}
\caption{BER v.s. transmit SNR, $N_t=128$, $K=8$, $n_{\max}=20$, 8-PSK}
\end{figure} 

In Fig. 6, we show the performance of different schemes for 8-PSK modulation with $N_t=128$ and $K=8$. For 1-bit quantized beamforming approaches, it is observed that the proposed optimization-based non-linear scheme achieves the best BER performance. For the symbol scaling approach, it is observed that in the case of 8-PSK, only a 1dB SNR loss is observed compared to the non-linear iterative `Pokemon' algorithm, and therefore the proposed low-complexity symbol scaling approach is more favourable in terms of the performance and complexity tradeoff.

\begin{figure}[!t]
\centering
\includegraphics[scale=0.45]{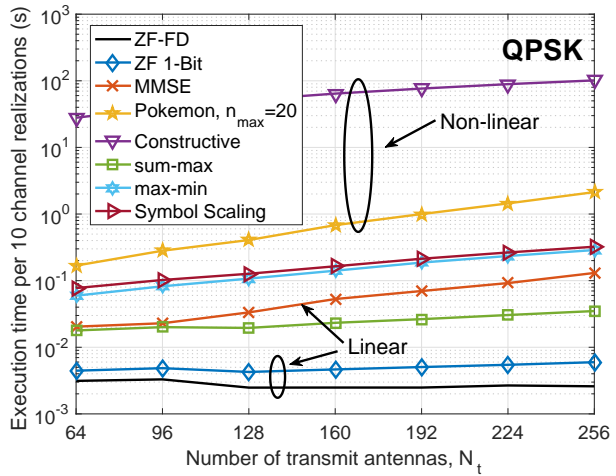}
\caption{Execution time for each scheme per 10 channel realizations, $K=4$, $n_{\max}=20$, QPSK}
\end{figure} 

In Fig. 7, we compare the computational complexity of each approach in terms of the execution time required per 10 channel realizations. It is not surprising to observe that the computational cost of the proposed non-linear scheme is the highest. Compared to the non-linear `Pokemon' algorithm, the execution time required for the proposed symbol scaling method is much less, especially for the `sum-max' case. For `Symbol Scaling' that returns the best performance based on Algorithm 5, the execution time required is similar to that of the `max-min', which validates our analysis in Section V-C that most of the computational cost in the allocation stage comes from the `Max-Min' criterion. Moreover, it is observed that the execution time of `Symbol Scaling' is approximately 12$\%$ of that of the `Pokemon' scheme in Fig. 7. This matches our analysis in (79) ($\frac{{{{\rm{C}}_{\rm{L}}}}}{{{{\rm{C}}_{\rm{P}}}}} \approx 0.12$ when $n_{\max}=20$), and the above complexity gains of the proposed symbol scaling approach therefore favour its practical application.

To further compare the proposed schemes with `Pokemon', in Fig. 8 we present the BER performance with different number of iterations for Pokemon. The number of iterations does not have an effect on other methods and therefore the BER for the other methods remains constant. It is observed that the performance of Pokemon improves as $n_{\rm max}$ increases. Nevertheless, we note that the improvement becomes less significant with a larger $n_{\rm max}$ and Pokemon achieves its best performance when $n_{\rm max}$ is around 25. An important observation is when $n_{\rm max}=2, 3$, where the computational cost of Pokemon and our proposed scheme is similar, as shown by (79), and our proposed symbol scaling approach is shown to achieve an improved performance, which validates the superiority of the proposed approach.

\begin{figure}[!t]
\centering
\includegraphics[scale=0.45]{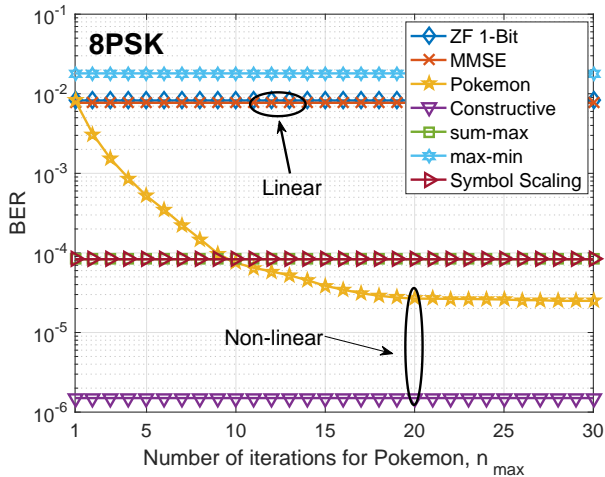}
\caption{BER v.s. Pokemon iteration number $n_{\rm max}$, $N_t=128$, $K=8$, $\rho=10$dB, 8-PSK}
\end{figure} 

To demonstrate the performance-complexity tradeoff directly, in Fig. 9 we depict the BER with respect to the number of floating-point operations required for a range of transmit antennas from $N_t=32$ to $N_t=128$, where the number of users is fixed as $K=8$. It can be observed that the proposed optimization-based method achieves the best performance at the cost of the highest complexity. An important comparison is between the proposed `Symbol Scaling' approach and the `Pokemon' scheme with $n_{\rm max}=2$, where we observe a significant performance gain of our proposed algorithm for the same computational complexity, especially when the number of antennas is large. Moreover, while the performance of the proposed low-complexity method based on `sum-max' achieves an inferior performance to the `Symbol Scaling' approach when $N_t$ is large, it indeed achieves a better BER performance with a lower computational cost compared to Pokemon with $n_{\rm max}=2$. Both of the above observations reveal the superiority of the proposed scheme based on symbol scaling.

\begin{figure}[!t]
\centering
\includegraphics[scale=0.45]{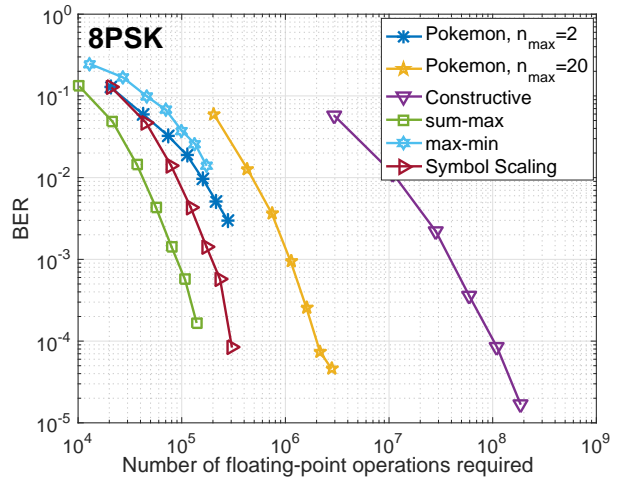}
\caption{BER v.s. analytical floating operations required, $K=8$, $\rho=10$dB, 8-PSK}
\end{figure} 

\begin{figure}[!t]
\centering
\includegraphics[scale=0.45]{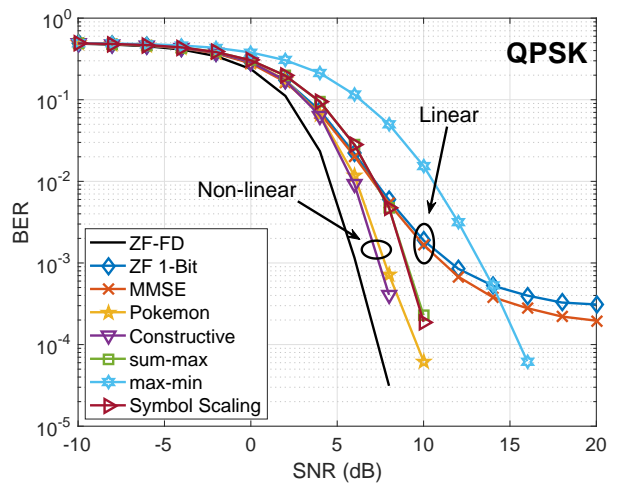}
\caption{BER v.s. transmit SNR, $N_t=128$, $K=16$, $n_{\max}=20$, QPSK, Imperfect CSI, $\beta=2.5$}
\end{figure} 

All the above results are based on the assumption of perfect CSI. In the following, we numerically investigate the performance of the proposed {approaches with imperfect CSI. The channel estimation techniques for massive MIMO with 1-bit quantization is an ongoing topic of research \cite{r16}, \cite{r25}, and an exact model for the imperfect CSI for this scenario is still not known. Therefore, in the following we employ a generic CSI model, where the BS only has knowledge of a noisy version of $\bf H$, given by
\begin{equation}
{\bf{\hat H}} = {\bf{H}} + {\bf{Q}}.
\end{equation}
In (80), ${\bf{\hat H}}$ is the obtained CSI at the BS. $\bf Q$ denotes an error matrix with ${\bf{Q}} \sim {\cal CN}\left( {{\bf{0}},\delta  \cdot {\bf{I}}} \right)$, where $\delta$ denotes the variance of the channel error. $\delta$ is modelled as inversely proportional to the transmit SNR and is expressed as $\delta  = {\beta  \mathord{\left/ {\vphantom {\beta  \rho }} \right. \kern-\nulldelimiterspace} \rho }$, where $\beta$ denotes the error coefficient \cite{r36}. The BER result with imperfect CSI is depicted in Fig. 10, where a similar trend can be observed. We can further observe that the proposed non-linear mapping method still achieves the best performance among the schemes with 1-bit quantization in the case of imperfect CSI, while the proposed low-complexity symbol scaling approach can achieve a comparable performance with a greatly reduced computational cost.

\section{Conclusion}
In this paper, we propose several transmit beamforming schemes for the massive MIMO downlink with 1-bit DACs based on the formulation of constructive interference, and we consider both a quantized linear method and a non-linear mapping approach. With the analysis of the Lagrangian and KKT conditions, the quantized linear scheme is mathematically proven to be equivalent to the quantized ZF beamforming. For the proposed non-linear mapping scheme, it is shown to be non-convex and solved by firstly relaxing the 1-bit quantization constraint, followed by a normalization. We further propose a low-complexity symbol scaling approach, where the quantized transmit signals are directly obtained. Numerical results reveal the superiority of the proposed symbol scaling scheme in small-scale MIMO systems. In the case of massive MIMO, the performance advantage of the proposed non-linear mapping method is validated, while the proposed symbol scaling scheme achieves a better performance-complexity tradeoff, which favours its usefulness in practical systems.

\appendix[Coordinate Transformation]
We employ 8-PSK modulation in Fig. 3 as the example to demonstrate the coordinate transformation, where we focus on the constellation point `A' in Fig. 3. Then, in the conventional real-imaginary complex plane, for node `B' in Fig. 3, we have 
\begin{equation}
\mathop {OB}\limits^ \to   = {{\bf{h}}_k}{{\bf{x}}_T} = {B_r} \cdot 1 + {B_i} \cdot j,
\end{equation}
where $1$ and $j$ are the bases, and we denote $\left( {{B_r},{B_i}} \right)$ as the corresponding coordinates. Based on (8), $B_r$ and $B_i$ are obtained as
\begin{equation}
\begin{aligned}
{B_r} & = \Re \left( {{{\bf{h}}_k}{{\bf{x}}_T}} \right) = \Re \left( {{{\bf{h}}_k}} \right){\bf{x}}_T^\Re - \Im \left( {{{\bf{h}}_k}} \right){\bf{x}}_T^\Im,\\
{B_i} & = \Im \left( {{{\bf{h}}_k}{{\bf{x}}_T}} \right) = \Im \left( {{{\bf{h}}_k}} \right){\bf{x}}_T^\Re + \Re \left( {{{\bf{h}}_k}} \right){\bf{x}}_T^\Im.
\end{aligned}
\end{equation}
In the plane expanded by the two detection thresholds that correspond to the constellation point `A', following (42) $\mathop {OB}\limits^ \to$ is decomposed into
\begin{equation}
\mathop {OB}\limits^ \to = {\bf h}_k{\bf x}_T = \alpha _k^\Re s_k^\Re  + \alpha _k^\Im s_k^\Im.
\end{equation}
Based on (41) and the fact that $\alpha _k^\Re$ and $\alpha _k^\Im$ are real numbers, (83) is further transformed into
\begin{equation}
\begin{aligned}
{\bf h}_k{\bf x}_T &= \alpha _k^\Re \left( {{a_k} + {b_k} \cdot j} \right) + \alpha _k^\Im \left( {{c_k} + {d_k} \cdot j} \right) \\
&= \left( {{a_k}\alpha _k^\Re  + {c_k}\alpha _k^\Im } \right) + \left( {{b_k}\alpha _k^\Re  + {d_k}\alpha _k^\Im } \right) \cdot j .
\end{aligned}
\end{equation}
By substituting (82) into (84), we obtain
\begin{equation}
\begin{aligned}
B_r & =  \Re \left( {{{\bf{h}}_k}} \right){\bf{x}}_T^\Re  - \Im \left( {{{\bf{h}}_k}} \right){\bf{x}}_T^\Im = {a_k}\alpha _k^\Re  + {c_k}\alpha _k^\Im, \\
B_i & =  \Im \left( {{{\bf{h}}_k}} \right){\bf{x}}_T^\Re  + \Re \left( {{{\bf{h}}_k}} \right){\bf{x}}_T^\Im = {b_k}\alpha _k^\Re  + {d_k}\alpha _k^\Im, 
\end{aligned}
\end{equation}
which leads to the expression of $\alpha _k^\Re$ and $\alpha _k^\Im$, given by
\begin{equation}
\begin{aligned}
&\alpha _k^\Re = \frac{{{d_k}{B_r} - {c_k}{B_i}}}{{{a_k}{d_k} - {b_k}{c_k}}}\\
=&\frac{{{d_k}\left[ {\Re \left( {{{\bf{h}}_k}} \right){\bf{x}}_T^\Re  - \Im \left( {{{\bf{h}}_k}} \right){\bf{x}}_T^\Im } \right] - {c_k}\left[ {\Im \left( {{{\bf{h}}_k}} \right){\bf{x}}_T^\Re  + \Re \left( {{{\bf{h}}_k}} \right){\bf{x}}_T^\Im } \right]}}{{{a_k}{d_k} - {b_k}{c_k}}} \\
 = &\frac{{{d_k}\Re \left( {{{\bf{h}}_k}} \right) - {c_k}\Im \left( {{{\bf{h}}_k}} \right)}}{{{a_k}{d_k} - {b_k}{c_k}}}{\bf{x}}_T^\Re  - \frac{{{d_k}\Im \left( {{{\bf{h}}_k}} \right) + {c_k}\Re \left( {{{\bf{h}}_k}} \right)}}{{{a_k}{d_k} - {b_k}{c_k}}}{\bf{x}}_T^\Im,
\end{aligned}
\end{equation}
and
\begin{equation}
\begin{aligned}
&\alpha _k^\Im = \frac{{{a_k}{B_i} - {b_k}{B_r}}}{{{a_k}{d_k} - {b_k}{c_k}}}\\
=&\frac{{{a_k}\left[ {\Im \left( {{{\bf{h}}_k}} \right){\bf{x}}_T^\Re  + \Re \left( {{{\bf{h}}_k}} \right){\bf{x}}_T^\Im } \right] - {b_k}\left[ {\Re \left( {{{\bf{h}}_k}} \right){\bf{x}}_T^\Re  - \Im \left( {{{\bf{h}}_k}} \right){\bf{x}}_T^\Im } \right]}}{{{a_k}{d_k} - {b_k}{c_k}}}\\
=&\frac{{{a_k}\Im \left( {{{\bf{h}}_k}} \right) - {b_k}\Re \left( {{{\bf{h}}_k}} \right)}}{{{a_k}{d_k} - {b_k}{c_k}}}{\bf{x}}_T^\Re  + \frac{{{a_k}\Re \left( {{{\bf{h}}_k}} \right) + {b_k}\Im \left( {{{\bf{h}}_k}} \right)}}{{{a_k}{d_k} - {b_k}{c_k}}}{\bf{x}}_T^\Im.
\end{aligned}
\end{equation}
The extension to the constellation points of other PSK modulations can be similarly obtained and is omitted for brevity.

\ifCLASSOPTIONcaptionsoff
  \newpage
\fi

\bibliographystyle{IEEEtran}
\bibliography{refs.bib}

% that's all folks
\end{document}